\documentclass[journal]{IEEEtran}

\usepackage[T1]{fontenc}

\usepackage{amsmath}
\usepackage[cmintegrals]{newtxmath}
\usepackage{bm}

\usepackage[acronym, nowarn]{glossaries}

\usepackage[caption=false,font=footnotesize,]{subfig}
\usepackage{tikz}
\usetikzlibrary{arrows}
\usetikzlibrary{calc}
\usetikzlibrary{positioning}
\usepackage{pgfplots}
\pgfplotsset{compat=newest}
\pgfplotsset{plot coordinates/math parser=true}

\usepackage[nolist]{acronym}
\usepackage{pgfplots}
\pgfplotsset{compat=newest}
\pgfplotsset{plot coordinates/math parser=false}

\usepackage{verbatim}

\usepackage{xspace}

\usepackage{siunitx}
\DeclareSIUnit\dBm{dBm}
\DeclareSIUnit\bits{bits}

\DeclareMathOperator{\E}{\mathbb{E}}

\renewcommand{\vec}[1]{\mathbf{#1}}  
\newcommand{\mat}[1]{\mathbf{#1}}
\newcommand{\tran}{^\mathsf{T}}
\newcommand{\herm}{^\mathsf{H}}
\newcommand{\dB}{\,\text{dB} }

\newcommand{\figref}[1]{Figure \ref{#1}}

\newcommand{\indcenBS}{\emph{\textcolor{sceOne}{single central BS}}\xspace}
\newcommand{\IndcenBS}{\emph{\textcolor{sceOne}{Single central BS}}\xspace}
\newcommand{\indtwoBS}{\emph{\textcolor{sceTwo}{two indoor BSs}}\xspace}
\newcommand{\IndtwoBS}{\emph{\textcolor{sceTwo}{Two indoor BSs}}\xspace}
\newcommand{\indfourBS}{\emph{\textcolor{sceSeven}{four indoor BSs}}\xspace}
\newcommand{\IndfourBS}{\emph{\textcolor{sceSeven}{Four indoor BSs}}\xspace}
\newcommand{\indfourtyBS}{\emph{\textcolor{sceNine}{fourty indoor BSs}}\xspace}
\newcommand{\IndfourtyBS}{\emph{\textcolor{sceNine}{Fourty indoor BSs}}\xspace}
\newcommand{\outBS}{\emph{\textcolor{sceTen}{outdoor BSs}}\xspace}
\newcommand{\OutBS}{\emph{\textcolor{sceTen}{Outdoor BSs}}\xspace}
\newcommand{\inoutBS}{\emph{\textcolor{sceEleven}{indoor-outdoor BSs}}\xspace}
\newcommand{\InoutBS}{\emph{\textcolor{sceEleven}{Indoor-outdoor BSs}}\xspace}

\newcommand{\legSceOne}{single central BS}
\newcommand{\legSceTwo}{two indoor BSs}
\newcommand{\legSceSeven}{four indoor BSs}
\newcommand{\legSceNine}{fourty indoor BSs}
\newcommand{\legSceTen}{outdoor BSs}
\newcommand{\legSceEleven}{indoor-outdoor BSs}

\newcommand{\mmimo}{massive MIMO\xspace}
\newcommand{\Mmimo}{Massive MIMO\xspace}
\newcommand{\nmimo}{network MIMO\xspace}
\newcommand{\Nmimo}{Network MIMO\xspace}

\newcommand{\locPre}{local precoding\xspace}
\newcommand{\LocPre}{Local precoding\xspace}
\newcommand{\cunit}{central processor\xspace}

\newcommand{\merWf}{mer\-cury\slash wa\-ter-fill\-ing\xspace}

\newcommand{\wf}{wa\-ter-fill\-ing\xspace}

\newacronym{ue}{UE}{user equipment}
\newacronym{bs}{BS}{base station}
\newacronym[user1={Zero-Forcing Beamforming}]{zfbf}{ZFBF}{zero-forcing beamforming}
\newacronym{siso}{SISO}{single-input single-output}
\newacronym{mimo}{MIMO}{mul\-ti\-ple-in\-put mul\-ti\-ple-out\-put}
\newacronym{miso}{MISO}{mul\-ti\-ple-in\-put sin\-gle-out\-put}
\newacronym[user1={Maximum Ratio Transmission}]{mrt}{MRT}{maximum ratio transmission}
\newacronym{bc}{BC}{broadcast channel}
\newacronym{mac}{MAC}{multiple-access channel}
\newacronym{los}{LOS}{line-of-sight}
\newacronym{nlos}{NLOS}{non line-of-sight}
\newacronym{ofdm}{OFDM}{orthogonal frequency-division multiplexing}
\newacronym{prb}{PRB}{physical resource block}
\newacronym{csi}{CSI}{channel state information}
\newacronym{cdf}{CDF}{cumulative distribution function}
\newacronym{ula}{ULA}{uniform linear array}
\newacronym{metis}{METIS}{Mobile and wireless communications Enablers for the Twenty-twenty Information Society}
\newacronym{eirp}{EIRP}{equivalent isotropically radiated power}
\newacronym{erp}{ERP}{effective radiated power}
\newacronym{iid}{i.i.d.\@}{independent and identically distributed}
\newacronym{awgn}{AWGN}{additive white Gaussian noise}
\newacronym{mmse}{MMSE}{minimum mean squared error}
\newacronym{qam}{QAM}{quadrature amplitude modulation}
\newacronym{ask}{ASK}{amplitude shift keying}
\newacronym{snr}{SNR}{signal-to-noise ratio}
\newacronym{sinr}{SINR}{signal-to-interference-plus-noise ratio}
\newacronym{csit}{CSIT}{channel state information at the transmitter}
\newacronym{dpc}{DPC}{dirty paper coding}
\newacronym{tdd}{TDD}{time division duplex}
\newacronym{fdd}{FDD}{frequency division duplex}
\newacronym{cs}{CS}{coordinated scheduling}
\newacronym{cb}{CB}{coordinated beamforming}
\newacronym[user1={Coordinated Scheduling/Coordinated Beamforming}]{cscb}{CS/CB}{coordinated scheduling/coordinated beamforming}
\newacronym[user1={Large-Scale MIMO}]{lsmimo}{LS-MIMO}{large-scale MIMO}
\newacronym{ia}{IA}{interference alignment}
\newacronym{svd}{SVD}{singular value decomposition}
\newacronym{wlog}{w.l.o.g.\@}{without loss of generality}
\newacronym{comp}{CoMP}{coordinated multipoint}
\newacronym{lte}{LTE}{Long Term Evolution}
\newacronym{ltea}{LTE-Advanced}{Long Term Evolution-Advanced}
\newacronym{5g}{5G}{5th generation mobile networks}
\newacronym{qos}{QoS}{quality-of-service}
\newacronym{quadriga}{QuaDRiGa}{Quasi Determinsitic Radio Channel Generator}
\newacronym{winner}{WINNER}{Wireless World Initiative New Radio}
\newacronym{winner2}{WINNER II}{Wireless World Initiative New Radio II}
\newacronym{winnerplus}{WINNER+}{Wireless World Initiative New Radio+}
\newacronym{3gpp}{3GGP}{3rd Generation Partnership Project}
\newacronym{isi}{ISI}{inter symbol interference}
\newacronym{fft}{FFT}{fast Fourier transform}
\newacronym{ifft}{IFFT}{inverse fast Fourier transform}
\newacronym{idft}{IDFT}{inverse discrete Fourier transform}
\newacronym[longplural={spectral efficiencies}]{se}{SE}{spectral efficiency}
\newacronym{nmse}{NMSE}{normalized mean squared error}
\newacronym{uwb}{UWB}{ultra wide band}
\newacronym{papr}{PAPR}{peak-to-average power ratio}
\newacronym{socp}{SOCP}{second order cone problem}
\newacronym{np}{NP}{non-deterministic polynomial-time}
\newacronym{wlan}{WLAN}{wireless local area network}
\newacronym{fcc}{FCC}{Federal Communications Commission}
\newacronym{cept}{CEPT}{European Conference of Postal and Telecommunications Administrations}
\newacronym{ecc}{ECC}{Electronic Communications Committee}
\newacronym{itu}{ITU}{International Telecommunication Union}
\newacronym{sar}{SAR}{specific absorption rate}
\newacronym{pzf}{P-ZF}{full-pilot zero-forcing}

\definecolor{sceOne}{HTML}{003359}
\definecolor{sceTwo}{HTML}{00778A}
\definecolor{sceSeven}{HTML}{679A1D}
\definecolor{sceNine}{HTML}{F9BA00}
\definecolor{sceTen}{HTML}{C4071B}
\definecolor{sceEleven}{HTML}{69085A}

\newcommand{\markSceOne}{x}
\newcommand{\markSceTwo}{triangle}
\newcommand{\markSceSeven}{square}
\newcommand{\markSceNine}{asterisk}
\newcommand{\markSceTen}{o}
\newcommand{\markSceEleven}{otimes}

\definecolor{mycolor1}{HTML}{E37222}%
\definecolor{mycolor2}{HTML}{003359}%
\definecolor{mycolor3}{HTML}{005293}%
\definecolor{mycolor4}{HTML}{0065BD}%
\definecolor{mycolor5}{HTML}{64A0C8}%
\definecolor{mycolor6}{HTML}{98C6EA}%

\newcommand{\legOffSet}{.6em} 
\newcommand{\legOffSetHalf}{.3em}

\begin{document}

\newlength{\plotheight}
\newlength{\plotwidth}
\setlength{\plotwidth}{.8\columnwidth}
\setlength{\plotheight}{.6\columnwidth}

\newlength{\mapheight}
\newlength{\mapwidth}
\setlength{\mapwidth}{.8\columnwidth}
\setlength{\mapheight}{.4\columnwidth}

\newlength{\halfmapwidth}
\newlength{\halfmapheight}
\setlength{\halfmapwidth}{.5\columnwidth}
\setlength{\halfmapheight}{.25\columnwidth}

%
\title{Analysis of Massive MIMO and Base Station Cooperation in an Indoor Scenario}
%
%
%

\author{Stefan~Dierks,
        Gerhard~Kramer,~\IEEEmembership{Fellow,~IEEE},
        Berthold~Panzner,~\IEEEmembership{Senior Member,~IEEE},
        and~Wolfgang~Zirwas,
\thanks{S. Dierks and G. Kramer are with the Chair for Communications Engineering, Technical University of Munich, Germany, e-mail: \{stefan.dierks, gerhard.kramer\}@tum.de}
\thanks{B. Panzner and W. Zirwas are with Nokia Networks, Munich, Germany, email: \{berthold.panzner, wolfgang.zirwas\}@nokia.com}
}
\maketitle

\begin{abstract}
The performance of centralized and distributed massive MIMO deployments are analyzed for indoor office scenarios.
The distributed deployments use one of the following precoding methods: (1) local precoding with local channel state information (CSI) to the user equipments (UEs) that it serves; (2) large-scale MIMO with local CSI to all UEs in the network; (3) network MIMO with global CSI. For the distributed deployments (2) and (3), it is shown that using twice as many base station antennas as data streams provides many of the massive MIMO benefits in terms of spectral efficiency and fairness. This is in contrast to the centralized deployment and the distributed deployment (1) where more antennas are needed.  Two of the main conclusions are that distributing base stations helps to overcome wall penetration loss; however, a backhaul is required to mitigate inter-cell interference.
The effect of estimation errors on the performance is also quantified.
\end{abstract}
%


%
\IEEEpeerreviewmaketitle

\section{Introduction}
\label{sec:Intro}

\IEEEPARstart{O}{ne} goal of new mobile radio communication standards, e.g., \gls{5g}, is to increase the \gls{se} per unit area or volume. For example, the \acrshort{metis} (\acrlong{metis}) project \cite{metis_d11} defines target traffic volume densities for different scenarios. 
One way to increase \gls{se} is by using \gls{mimo} schemes. \Gls{mimo} allows one node to transmit several streams to one or more \glspl{ue} using spatial degrees-of-freedom.

\emph{Massive \gls{mimo}} refers to a ``vast'' over-provisioning of \gls{bs} antennas as compared to the number of served single antenna \glspl{ue} \cite{Marzetta_mMIMO2010}. 
\Mmimo is also known as ``Very Large \gls{mimo}'', ``Hyper \gls{mimo}'', ``Full Dimension \gls{mimo}'', ``Large-Scale Antenna Systems'', or ``ARGOS'' \cite{Larsson2014}.
However, the term \mmimo is not clearly defined. Massive \gls{mimo} may refer to any \gls{mimo} configuration beyond the largest \gls{mimo} mode in the current LTE standard (at present 8x8), e.g., 100 antennas or more \cite{Rusek13}, or it may simply refer to a ``large`` number of antennas at the \glspl{bs}. A more precise definition of massive \gls{mimo} is based on the ratio $M/K$ of serving \gls{bs} antennas $M$ to the number $K$ of active \glspl{ue}. However, the ratio $M/K$ for which one can speak of \mmimo depends on the performance metric, the scenario, etc. \cite{Bjoernson16}.

\Mmimo claims several advantages over conventional \gls{mimo} \cite{Larsson2014}:
\begin{itemize}
 \item \emph{\Mmimo increases capacity by 10 times or more and simultaneously increases energy efficiency.} The transmit signals are directed precisely to the \glspl{ue} through precoding which reduces interference. Each additional antenna increases the precoding degrees-of-freedom assuming no mutual coupling and a sufficiently complex propagation environment \cite{Rusek13}.
 \item \emph{Inexpensive, low-power components suffice.} A large number of \gls{bs} antennas makes the system robust against noise, fading and hardware impairments or even failure of antenna elements. This allows simpler transmitters and receivers at the \gls{bs}, e.g., few or one bit quantization, hybrid digital-analog precoding, and constant envelope precoding \cite{Bjoernson16}.
 \item \emph{Precoding simplifies.} Simple linear precoding has a vanishing gap to optimal precoding \cite{Marzetta_mMIMO2010,Rusek13,Bjoernson16}. For instance, the performance gap between linear \gls{zfbf} \cite{bjoernson14} and the optimal, non-linear \gls{dpc} \cite{Costa83} vanishes with an increasing number of \gls{bs} antennas. \Gls{mrt} is also asymptotically optimal as the number of \gls{bs} antennas increases, but for a smaller number of \gls{bs} antennas \gls{mrt} performs well only in the low \gls{snr} regime \cite{Rusek13}.
 \item \emph{The multiple-access layer simplifies.} The channel hardens by the law of large numbers \cite{Larsson2014,Rusek13}. This means that all subcarriers experience similar small-scale fading and the \gls{ue} channel vectors become orthogonal. Hence scheduling does not improve performance because all \glspl{ue} can be active on all subcarriers. Only power control is needed to distribute the power depending on the slowly varying large-scale fading \cite{Bjoernson16}.
 \item \emph{The latency is reduced.} Since all \glspl{ue} can always be active, \glspl{ue} need not wait for good fading conditions.
 \item \emph{\Mmimo is robust to jamming and interference.} The surplus of precoding degrees-of-freedom can be used to cancel interference or jamming.
\end{itemize}

Most massive \gls{mimo} studies consider wide area outdoor scenarios \cite{Marzetta_mMIMO2010,Larsson2014,Bjoernson16}.
However, most mobile traffic is generated by indoor users \cite{Femtocells.2010}.
We analyze the performance of different \gls{bs} deployments with different levels of cooperation for the \gls{3gpp} indoor office scenario \cite{3GPP_TR36_814.2010}. Our approach is as follows. We fix the number of active, single antenna \glspl{ue} and sweep the ratio of total number of \gls{bs} antennas to the number of active \glspl{ue} from one to ten. We find that a ratio of twice as many \gls{bs} antennas provides most of the massive \gls{mimo} benefits. We further find that this ratio is a good tradeoff between number of antennas versus \gls{se}. We present suboptimal transmission schemes that approach a capacity upper bound. We analyze fairness using Jain's index \cite{jainsindex}.

We further compare massive \gls{mimo} to distributed \gls{mimo}. The motivation is that placing a single massive \gls{mimo} \gls{bs} at the center of a building causes \glspl{ue} to experience large path loss and high wall penetration loss. We compare this deployment to distributed \glspl{bs}. We find that distributed indoor \glspl{bs} with cooperation achieve a substantial performance gain at the cost of a backhaul connection, while the gain achieved with cooperation between outdoor \glspl{bs} and a single indoor \gls{bs} is smaller.
With increasing capability of the backhaul, the cooperation level can be increased which allows to achieve the same performance with fewer \gls{bs} antennas.

Finally, we quantify the performance loss due to channel estimation error.
Like in conventional \gls{mimo}, \gls{csi} is required to enable precoding. Acquiring \gls{csi} might be more difficult in \mmimo due to the many antennas. \Gls{fdd} requires a 
pilot sequence for each \gls{bs} antenna, while \gls{tdd} suffers from pilot contamination \cite{Marzetta_mMIMO2010}.

Our results help guide design choices for future mobile radio communication systems, e.g., \gls{ltea} and \gls{5g}.
We presented preliminary results in \cite{ofdmWorkshop,globeComMMIMOWS,eusipco}, and we add the following results.
\begin{itemize}
 \item Instead of using \wf to allocate power, we use \merWf, which is optimal for finite modulation alphabets \cite{LozanoJul06}.
 \item We analyze two additional deployments (the \indtwoBS deployment and the \indfourtyBS deployment).
 \item We add \gls{lsmimo} as an example of an interference coordination scheme.
 \item We analyze fairness for Gaussian modulation.
\end{itemize}

We denote vectors with bold lower case letters, and matrices with bold upper case letters. The transpose of $\mat{X}$ is $\mat{X}\tran$ and the complex conjugate transpose is $\mat{X}\herm$. The Euclidean norm of $\vec{x}$ is $\left\| \vec{x} \right\|_2$, and the Frobenius norm of $\mat{X}$ is $\left\| \mat{X} \right\|_F$. We denote a diagonal matrix having diagonal entries $\vec{x}$ as $\text{diag}\left(\vec{x}\right)$.

\section{System Model}
\label{sec:sys_mod}
Consider the downlink in the \gls{3gpp} ``A1 - Indoor Office'' scenario in the \gls{winner2} deliverable \cite{IST-WINNERIIDeliverable1.1.22008a}, see Fig. \ref{fig:deployments}. The \glspl{ue} are served by \glspl{bs} located inside and outside the building, and we describe the \gls{bs} deployments in Section \ref{sec:bsdeployments}.
We consider $K$ single antenna \glspl{ue} and \gls{ofdm}. For each subcarrier we obtain a \gls{bc}. The received signal of the $k$-th \gls{ue} for one subcarrier is
\begin{equation}
\label{eq:oneUErec}
y_{k} = \vec{h}_{k}\herm \vec{x} + z_k \quad k\in\left\{1,\ldots,K\right\}
\end{equation}
where $\vec{h}_k\herm = [\vec{h}_{k,1}\herm, \ldots, \vec{h}_{k,N_\text{BS}}\herm ]$ is the vector of channel coefficients from all $N_\text{BS}$ \glspl{bs} to the $k$-th \gls{ue}. The $i$-th \gls{bs} has $M_i$ \gls{bs} antennas with the channel coefficients $\vec{h}_{k,i}\herm$. The dimension of $\vec{h}_k\herm$ is $M=\sum_{i=1}^{N_\text{BS}} M_i$.
The transmit signal vectors are collected in $\vec{x} = [\vec{x}_{1}\tran, \ldots, \vec{x}_{N_\text{BS}}\tran ]\tran$ and the $z_1,z_2,\ldots,z_K$ are independent proper complex \gls{awgn} random variables with variance $\sigma_N^2$. 
The received signals of all \glspl{ue} $\vec{y} = [y_1,\ldots,y_K]\tran$ for one subcarrier are collected in the vector
\begin{equation}
\label{eq:allUErec}
\vec{y} = \mat{H}\herm \vec{x} + \vec{z}
\end{equation}
where $\mat{H}\herm = [\vec{h}_1\herm,\ldots,\vec{h}_K\herm]$ and $\vec{z} = [z_1,\ldots,z_K]\tran$.

For linear precoding the transmit signals vector $\vec{x}$ is
\begin{equation}
\label{eq:precoding}
\vec{x} = \mat{W} \vec{s}
\end{equation}
where $\mat{W} = [\vec{w}_1,\ldots,\vec{w}_K]$ is the matrix of the precoding vectors and $\vec{s} = [s_1,\ldots,s_K]\tran$ is the vector of transmit symbols. We assume that $\E\left[\left|s_k\right|^2\right]=1$ for $k\in\left\{1,2,\ldots,K\right\}$.
We consider per-\gls{bs} sum-power constraints
\begin{equation}
\label{eq:pwConst}
\sum_{f=1}^{N_\text{SC}} \E\left[\left\|\vec{x}_{i}^{(f)}\right\|^2_2\right] = \sum_{f=1}^{N_\text{SC}}\left\|\mat{W}^{(f)}_i \right\|^2_F \leq P_i \quad\forall i \in \left\{1,\ldots, N_\text{BS}\right\}
\end{equation}
where $N_\text{SC}$ is the number of subcarriers, and $\mat{W}^{(f)}_i$ is the part of the precoding matrix that creates the transmit signals at the $i$-th \gls{bs} $\vec{x}^{(f)}_i$.
We often omit the subcarrier index if $f$ is clear from the context.

\section{Transmission Schemes}
\label{sec:trans_schem}

Interference management is important for modern wireless communication standards like \gls{lte} \cite{Boudreau09}, \gls{ltea} \cite{Lee12}, and for future standards like \gls{5g}.
A general framework and optimization algorithms for multi-cell scenarios with different levels of cooperation are presented in \cite{CIT069}.  
We are interested in transmission schemes with low complexity. For ease of notation we describe the principle for a single subcarrier and omit the subcarrier index.

\subsection{Local Precoding}
\emph{\LocPre} \glspl{bs} determine the transmit signals and the scheduled \glspl{ue} locally. They treat inter-cell interference as noise and thus interference limits reliable transmission in many scenarios. As a result, backhaul requirements are low and only local \gls{csi} is required.

Suppose each \gls{ue} is served by the \gls{bs} with the maximum average \gls{snr}.
We use \gls{zfbf} \cite{Wiesel2008} to mitigate intra-cell interference.
The local precoding matrix at the $i$-th \gls{bs} is
\begin{equation}
 \label{eq:localZFBF}
 \mat{W}_{i} = \mat{H}_{i,i}\left(\mat{H}_{i,i}\herm\mat{H}_{i,i}\right)^{-1} \text{diag}\left(\tilde{\vec{p}}_i\right)^{\frac{1}{2}}
\end{equation}
where $\mat{H}_{i,i}\herm$ is the channel matrix from the $i$-th \gls{bs} to its $K_i$ \glspl{ue}, and $\tilde{\vec{p}}_i$ is the power allocation vector at the $i$-th \gls{bs}.
\Gls{zfbf} requires that the $i$-th \gls{bs} serves at most $M_i$ \glspl{ue}, i.e., we have $K_i\leq M_i$. If $K_i>M_i$ then we use the low complexity scheduling algorithm from \cite{Zhang2008} to select $M_i$ \glspl{ue}. Note that the set of scheduled \glspl{ue} may be different on each subcarrier.
We use \merWf \cite{LozanoJul06} at each \gls{bs} to allocate power according to a per-\gls{bs} power constraint.

\subsection{Large-Scale MIMO}
Interference coordination has each \gls{bs} estimate its channels to all \glspl{ue}, and each \gls{bs} exchange its \gls{csi} with the other \glspl{bs}. The resulting global \gls{csi} lets us coordinate the transmissions of the \glspl{bs}, e.g., by power allocation, precoding, and scheduling.
In contrast, for \locPre each \gls{bs} estimates only its channels to the \glspl{ue} it serves.
Interference coordination has each \gls{ue} served by a single \gls{bs}, and the backhaul requirements are modest because a signal-level synchronization of the \glspl{bs} is not needed \cite{Gesbert10}.
The \glspl{ue} served by each \gls{bs} are determined based on maximal \gls{snr} as for \locPre. 
The coordination can be accomplished at a \cunit or locally at the \glspl{bs}. The distributed, local coordination can be realized in a competitive (game theoretic) way or with the help of control messages over the backhaul. Note that to reduce the backhaul requirements some coordination schemes exchange little \gls{csi}, and some schemes exchange control messages instead.

The coordination schemes can be categorized as follows \cite{Gesbert10}:
\begin{itemize}
 \item \Gls{cs} has the scheduling and power allocation optimized jointly by all \glspl{bs}. 
 \item \Gls{cb} has the precoding coordinated using available precoding degrees-of-freedom to reduce interference. 
 \item The combination of \gls{cs} and \gls{cb}, which is called \gls{cscb}, is more common than pure \gls{cb}. 
 \item For \gls{cs}, \gls{cb}, and \gls{cscb} interference is treated as noise. Performance can improve if interference is detected at the \glspl{ue} \cite{Gesbert10}. Interference detection can be supported by coding at the transmitter, e.g., by interference alignment.
\end{itemize} 

We consider \emph{\acrfull{lsmimo}} \cite{Hosseini14} as an example of an interference coordination scheme. \Gls{lsmimo} is a linear \gls{cb} scheme which does not exchange \gls{csi} or control messages over the backhaul.
With sufficiently many antennas, each \gls{bs} uses \gls{zfbf} to mitigate the interference created at all \glspl{ue} and thereby creates parallel interference-free channels to the \glspl{ue} it serves.
This is feasible only if the number $M_i$ of antennas at the $i$-th \gls{bs} is at least as large as the total number of \glspl{ue}, i.e., we have $M_i\geq K$.
Hence \gls{lsmimo} is feasible only if $M \geq N_\text{BS}K$.

Note that \gls{lsmimo} can be made feasible by scheduling a subset of \glspl{ue}. However, we analyze \gls{lsmimo} only if $M \geq N_\text{BS}K$.
We use \gls{zfbf}
\begin{equation}
 \mat{W}_i = \mat{H}_i\left(\mat{H}_i\herm\mat{H}_i\right)^{-1} \text{diag}\left(\tilde{\vec{p}}_i\right)^{\frac{1}{2}}
\end{equation}
where $\mat{H}_i\herm$ is the channel matrix from the $i$-th \gls{bs} to all \glspl{ue}, and the power is allocated by \merWf \cite{LozanoJul06}.
\Mmimo approaches the zero-forcing behavior of \gls{lsmimo} with increasing $M_i$ because channels to the \glspl{ue} of the other \glspl{bs} become orthogonal to the channels of the served \glspl{ue} \cite{Hosseini14}.

\subsection{Network MIMO}
\label{sec:nmimo}
\Nmimo requires that the \glspl{bs} are connected by a backhaul with low delay and high throughput, and that the \glspl{bs} are synchronized. The distributed \glspl{bs} act as one \gls{bs} with distributed antennas, and the downlink channel becomes a \gls{bc}. In contrast to interference coordination, \nmimo may have interference enhance the signals at the \glspl{ue}.
\Nmimo can be realized by a \cunit or by exchanging messages between the \glspl{bs}.

For our \emph{\nmimo} scheme, we assume a perfect backhaul with unlimited capacity and zero delay. We let all \glspl{bs} act as a single \gls{bs} with distributed antennas and apply \gls{zfbf} with per-\gls{bs} power constraints. The classic \gls{mac}-\gls{bc} duality does not determine the optimal precoder for per-\gls{bs} power constraints \cite{Gesbert10}.
We use a low-complexity and suboptimal approach and determine for each subcarrier the \gls{zfbf} precoding matrix
\begin{equation}
\label{eq:ZFBFopt}
\mat{W} = \mat{H}\left(\mat{H}\herm\mat{H}\right)^{-1} \text{diag}\left(\tilde{\vec{p}}\right)^{\frac{1}{2}}
\end{equation}
where $\tilde{\vec{p}}$ is the power allocation vector.
We use \merWf \cite{LozanoJul06} to allocate power according to a total power constraint 
\begin{equation}
\sum_{f=1}^{N_\text{SC}}\E\left[\left\|\vec{x}^{(f)}\right\|^2_2\right] \leq \sum_{i=1}^{N_\text{BS}} P_i.
\end{equation}
Next, we determine the transmit power at each \gls{bs} and scale the precoding matrix $\mat{W}$ so that the per-\gls{bs} power constraint is satisfied at the \gls{bs} with the maximal transmit power. Note that the other \glspl{bs} could transmit with higher power. Hence this is a suboptimal approach, and better approaches can be found, e.g., see \cite{Zhang10,Shi08}.

\Nmimo helps to avoid rank deficient and poorly conditioned channel matrices which are caused by spatial correlations or by the ``keyhole'' effect \cite{zhang2004capacity}.
\Nmimo is sometimes called ``distributed \gls{mimo}'', ``\gls{mimo} cooperation'', ``coherently coordinated transmission'', ``Joint Processing CoMP'', ``Joint Transmission CoMP'', ``C-RAN (Cloud-RAN)'' or ``p-cell'' \cite{pcell15}.

\section{Indoor Scenario}
\begin{figure}[t!]
  \centering
%
%
\begin{tikzpicture}

\begin{axis}[%
width=\mapwidth,
height=1.6\mapheight,
scale only axis,
point meta min=5,
point meta max=95,
xmin=-50,
xmax=50,
xtick={-40, -20,   0,  20,  40},
xticklabels={$\SI{-40}{m}$, $\SI{-20}{m}$, $\SI{0}{m}$, $\SI{20}{m}$, $\SI{40}{m}$},
ymin=-40,
ymax=40,
ytick={-40,-20, 0, 20,40},
yticklabels={$\SI{-40}{m}$, $\SI{-20}{m}$, $\SI{0}{m}$, $\SI{20}{m}$, $\SI{40}{m}$},
axis background/.style={fill=white},
axis x line*=bottom,
axis y line*=left,
]
\input{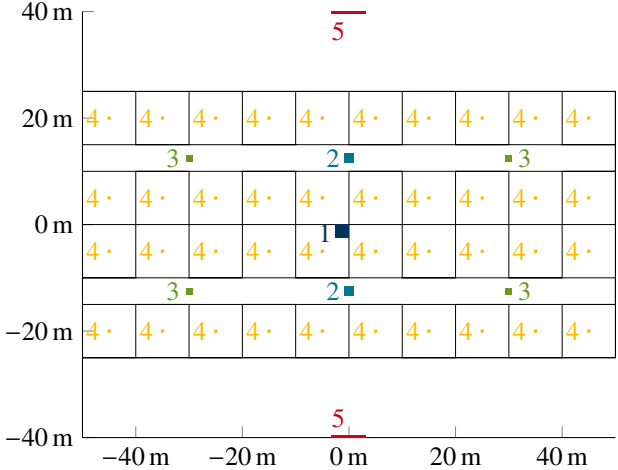}

\addplot [draw=none,fill=sceOne,forget plot]
  table[row sep=crcr]{%
0	0\\
0	-2.5\\
-2.5	-2.5\\
-2.5	0\\
0	0\\
};
\path[color=sceOne] (axis cs: -1.5,-1.5) node[left] {1};

\addplot [draw=none,fill=sceTwo,forget plot]
  table[row sep=crcr]{%
-.9	11.6\\
-.9	13.4\\
.9	13.4\\
.9	11.6\\
-.9	11.6\\
};
\path[color=sceTwo] (axis cs: 0,12.5) node[left] {2};
\addplot [draw=none,fill=sceTwo,forget plot]
  table[row sep=crcr]{%
-.9	-11.6\\
-.9	-13.4\\
.9	-13.4\\
.9	-11.6\\
-.9	-11.6\\
};
\path[color=sceTwo] (axis cs: 0,-12.5) node[left] {2};

\addplot [draw=none,fill=sceSeven,forget plot]
  table[row sep=crcr]{%
-30.63	11.87\\
-30.63	13.13\\
-29.37	13.13\\
-29.37	11.87\\
-30.63	11.87\\
};
\path[color=sceSeven] (axis cs: -30,12.5) node[left] {3};
\addplot [draw=none,fill=sceSeven,forget plot]
  table[row sep=crcr]{%
30.63	-11.87\\
30.63	-13.13\\
29.37	-13.13\\
29.37	-11.87\\
30.63	-11.87\\
};
\path[color=sceSeven] (axis cs: -30,-12.5) node[left] {3};
\addplot [draw=none,fill=sceSeven,forget plot]
  table[row sep=crcr]{%
-30.63	-11.87\\
-30.63	-13.13\\
-29.37	-13.13\\
-29.37	-11.87\\
-30.63	-11.87\\
};
\path[color=sceSeven] (axis cs: 30,-12.5) node[right] {3};
\addplot [draw=none,fill=sceSeven,forget plot]
  table[row sep=crcr]{%
30.63	11.87\\
30.63	13.13\\
29.37	13.13\\
29.37	11.87\\
30.63	11.87\\
};
\path[color=sceSeven] (axis cs: 30,12.5) node[right] {3};


\path[color=sceNine] (axis cs: -45,-20) node[left] {4};
\filldraw[fill=sceNine, draw=sceNine] (axis cs: -45.2,-20.2) rectangle (axis cs: -44.8,-19.8);
\path[color=sceNine] (axis cs: -35,-20) node[left] {4};
\filldraw[fill=sceNine, draw=sceNine] (axis cs: -35.2,-20.2) rectangle (axis cs: -34.8,-19.8);
\path[color=sceNine] (axis cs: -25,-20) node[left] {4};
\filldraw[fill=sceNine, draw=sceNine] (axis cs: -25.2,-20.2) rectangle (axis cs: -24.8,-19.8);
\path[color=sceNine] (axis cs: -15,-20) node[left] {4};
\filldraw[fill=sceNine, draw=sceNine] (axis cs: -15.2,-20.2) rectangle (axis cs: -14.8,-19.8);
\path[color=sceNine] (axis cs: -5,-20) node[left] {4};
\filldraw[fill=sceNine, draw=sceNine] (axis cs: -5.2,-20.2) rectangle (axis cs: -4.8,-19.8);
\path[color=sceNine] (axis cs: 5,-20) node[left] {4};
\filldraw[fill=sceNine, draw=sceNine] (axis cs: 4.8,-20.2) rectangle (axis cs: 5.2,-19.8);
\path[color=sceNine] (axis cs: 15,-20) node[left] {4};
\filldraw[fill=sceNine, draw=sceNine] (axis cs: 14.8,-20.2) rectangle (axis cs: 15.2,-19.8);
\path[color=sceNine] (axis cs: 25,-20) node[left] {4};
\filldraw[fill=sceNine, draw=sceNine] (axis cs: 24.8,-20.2) rectangle (axis cs: 25.2,-19.8);
\path[color=sceNine] (axis cs: 35,-20) node[left] {4};
\filldraw[fill=sceNine, draw=sceNine] (axis cs: 34.8,-20.2) rectangle (axis cs: 35.2,-19.8);
\path[color=sceNine] (axis cs: 45,-20) node[left] {4};
\filldraw[fill=sceNine, draw=sceNine] (axis cs: 44.8,-20.2) rectangle (axis cs: 45.2,-19.8);
\path[color=sceNine] (axis cs: -45,-5) node[left] {4};
\filldraw[fill=sceNine, draw=sceNine] (axis cs: -45.2,-5.2) rectangle (axis cs: -44.8,-4.8);
\path[color=sceNine] (axis cs: -35,-5) node[left] {4};
\filldraw[fill=sceNine, draw=sceNine] (axis cs: -35.2,-5.2) rectangle (axis cs: -34.8,-4.8);
\path[color=sceNine] (axis cs: -25,-5) node[left] {4};
\filldraw[fill=sceNine, draw=sceNine] (axis cs: -25.2,-5.2) rectangle (axis cs: -24.8,-4.8);
\path[color=sceNine] (axis cs: -15,-5) node[left] {4};
\filldraw[fill=sceNine, draw=sceNine] (axis cs: -15.2,-5.2) rectangle (axis cs: -14.8,-4.8);
\path[color=sceNine] (axis cs: -5,-5) node[left] {4};
\filldraw[fill=sceNine, draw=sceNine] (axis cs: -5.2,-5.2) rectangle (axis cs: -4.8,-4.8);
\path[color=sceNine] (axis cs: 5,-5) node[left] {4};
\filldraw[fill=sceNine, draw=sceNine] (axis cs: 4.8,-5.2) rectangle (axis cs: 5.2,-4.8);
\path[color=sceNine] (axis cs: 15,-5) node[left] {4};
\filldraw[fill=sceNine, draw=sceNine] (axis cs: 14.8,-5.2) rectangle (axis cs: 15.2,-4.8);
\path[color=sceNine] (axis cs: 25,-5) node[left] {4};
\filldraw[fill=sceNine, draw=sceNine] (axis cs: 24.8,-5.2) rectangle (axis cs: 25.2,-4.8);
\path[color=sceNine] (axis cs: 35,-5) node[left] {4};
\filldraw[fill=sceNine, draw=sceNine] (axis cs: 34.8,-5.2) rectangle (axis cs: 35.2,-4.8);
\path[color=sceNine] (axis cs: 45,-5) node[left] {4};
\filldraw[fill=sceNine, draw=sceNine] (axis cs: 44.8,-5.2) rectangle (axis cs: 45.2,-4.8);
\path[color=sceNine] (axis cs: -45,5) node[left] {4};
\filldraw[fill=sceNine, draw=sceNine] (axis cs: -45.2,5.2) rectangle (axis cs: -44.8,4.8);
\path[color=sceNine] (axis cs: -35,5) node[left] {4};
\filldraw[fill=sceNine, draw=sceNine] (axis cs: -35.2,5.2) rectangle (axis cs: -34.8,4.8);
\path[color=sceNine] (axis cs: -25,5) node[left] {4};
\filldraw[fill=sceNine, draw=sceNine] (axis cs: -25.2,5.2) rectangle (axis cs: -24.8,4.8);
\path[color=sceNine] (axis cs: -15,5) node[left] {4};
\filldraw[fill=sceNine, draw=sceNine] (axis cs: -15.2,5.2) rectangle (axis cs: -14.8,4.8);
\path[color=sceNine] (axis cs: -5,5) node[left] {4};
\filldraw[fill=sceNine, draw=sceNine] (axis cs: -5.2,5.2) rectangle (axis cs: -4.8,4.8);
\path[color=sceNine] (axis cs: 5,5) node[left] {4};
\filldraw[fill=sceNine, draw=sceNine] (axis cs: 4.8,5.2) rectangle (axis cs: 5.2,4.8);
\path[color=sceNine] (axis cs: 15,5) node[left] {4};
\filldraw[fill=sceNine, draw=sceNine] (axis cs: 14.8,5.2) rectangle (axis cs: 15.2,4.8);
\path[color=sceNine] (axis cs: 25,5) node[left] {4};
\filldraw[fill=sceNine, draw=sceNine] (axis cs: 24.8,5.2) rectangle (axis cs: 25.2,4.8);
\path[color=sceNine] (axis cs: 35,5) node[left] {4};
\filldraw[fill=sceNine, draw=sceNine] (axis cs: 34.8,5.2) rectangle (axis cs: 35.2,4.8);
\path[color=sceNine] (axis cs: 45,5) node[left] {4};
\filldraw[fill=sceNine, draw=sceNine] (axis cs: 44.8,5.2) rectangle (axis cs: 45.2,4.8);
\path[color=sceNine] (axis cs: -45,20) node[left] {4};
\filldraw[fill=sceNine, draw=sceNine] (axis cs: -45.2,20.2) rectangle (axis cs: -44.8,19.8);
\path[color=sceNine] (axis cs: -35,20) node[left] {4};
\filldraw[fill=sceNine, draw=sceNine] (axis cs: -35.2,20.2) rectangle (axis cs: -34.8,19.8);
\path[color=sceNine] (axis cs: -25,20) node[left] {4};
\filldraw[fill=sceNine, draw=sceNine] (axis cs: -25.2,20.2) rectangle (axis cs: -24.8,19.8);
\path[color=sceNine] (axis cs: -15,20) node[left] {4};
\filldraw[fill=sceNine, draw=sceNine] (axis cs: -15.2,20.2) rectangle (axis cs: -14.8,19.8);
\path[color=sceNine] (axis cs: -5,20) node[left] {4};
\filldraw[fill=sceNine, draw=sceNine] (axis cs: -5.2,20.2) rectangle (axis cs: -4.8,19.8);
\path[color=sceNine] (axis cs: 5,20) node[left] {4};
\filldraw[fill=sceNine, draw=sceNine] (axis cs: 4.8,20.2) rectangle (axis cs: 5.2,19.8);
\path[color=sceNine] (axis cs: 15,20) node[left] {4};
\filldraw[fill=sceNine, draw=sceNine] (axis cs: 14.8,20.2) rectangle (axis cs: 15.2,19.8);
\path[color=sceNine] (axis cs: 25,20) node[left] {4};
\filldraw[fill=sceNine, draw=sceNine] (axis cs: 24.8,20.2) rectangle (axis cs: 25.2,19.8);
\path[color=sceNine] (axis cs: 35,20) node[left] {4};
\filldraw[fill=sceNine, draw=sceNine] (axis cs: 34.8,20.2) rectangle (axis cs: 35.2,19.8);
\path[color=sceNine] (axis cs: 45,20) node[left] {4};
\filldraw[fill=sceNine, draw=sceNine] (axis cs: 44.8,20.2) rectangle (axis cs: 45.2,19.8);


\addplot [draw=none,fill=sceTen,forget plot]
  table[row sep=crcr]{%
3.2	40\\
3.2	39.5\\
-3.2	39.5\\
-3.2	40\\
3.2	40\\
};
\path[color=sceTen] (axis cs: -2,39.75) node[ below] {5};
\addplot [draw=none,fill=sceTen,forget plot]
  table[row sep=crcr]{%
3.2	-40\\
3.2	-39.5\\
-3.2	-39.5\\
-3.2	-40\\
3.2	-40\\
};
\path[color=sceTen] (axis cs: -2,-39.75) node[above] {5};



\end{axis}
\end{tikzpicture}%
  \caption{Base station deployments in the indoor office scenario \cite{IST-WINNERIIDeliverable1.1.22008a}.}
  \label{fig:deployments}
\end{figure}

\figref{fig:deployments} shows the layout of the indoor office scenario defined as ``A1 - Indoor Office'' in the \gls{winner2} deliverable D.1.1.2 \cite{IST-WINNERIIDeliverable1.1.22008a}. The \glspl{ue} are located \SI{1.5}{\meter} above the floor inside the building.
We use the \gls{quadriga} \cite{quadriga} to generate channel coefficients.

The indoor channels are generated according to the ``A1 - Indoor Office'' channel model parameters \cite{IST-WINNERIIDeliverable1.1.22008a}. There are two parameter sets for \gls{los} and for \gls{nlos} conditions.
For \gls{nlos} conditions a wall penetration loss is added, where the wall penetration loss is determined by counting the number of walls between each \gls{bs} and \gls{ue} beyond the first penetrated wall. When counting the number of walls, paths along the corridors are considered as alternatives to the direct path, which might penetrate more walls.

The outdoor-to-indoor channels are generated according to the ``B4 - Outdoor to indoor'' channel model parameters defined in \gls{winnerplus} deliverable D5.3 \cite{winnerPlusD5.3}. The outdoor \glspl{bs} are below rooftop micro \glspl{bs}. We assume a \gls{los} path from the \gls{bs} to the outside wall of the building. For each \gls{ue} the pathloss is calculated based on the path through the point on an outside wall nearest to the \gls{ue}.
The number of penetrated walls is determined as for the indoor \glspl{bs}.

\subsection{Base Station Deployments}
\label{sec:bsdeployments}
We define six different \gls{bs} deployments which are shown in \figref{fig:deployments}.
\begin{itemize}
 \item 
 \IndcenBS
 is a single \gls{bs} with $M$ antennas located in the corner of the room southwest of the center (``\textcolor{sceOne}{1}'' in \figref{fig:deployments}). This is a classical \mmimo deployment.
 \item 
 \IndtwoBS
 are two \glspl{bs} with $M/2$ antennas each. One \gls{bs} is located in the center of each corridor (``\textcolor{sceTwo}{2}'').
 \item
 \IndfourBS
 are four \glspl{bs} with $M/4$ antennas each. Two \glspl{bs} are located in each corridor (``\textcolor{sceSeven}{3}'').
 \item
 \IndfourtyBS
 are forty \glspl{bs} with $M/40$ antennas each. One \gls{bs} is located in the center of each room (``\textcolor{sceNine}{4}''). This is similar to the deployment of p-cell \cite{pcell15}.
 \item
 \OutBS
 are two \glspl{bs} with $M/2$ antennas each. They are located \SI{15}{\meter} north/south of the middle of the north/south outside wall (``\textcolor{sceTen}{5}'').
 \item
 \InoutBS
 are three \glspl{bs} with $M/3$ antennas each. One \gls{bs} is in the location of the \indcenBS deployment (``\textcolor{sceOne}{1}'') while two are in the location of the \outBS deployment (``\textcolor{sceTen}{5}'').
\end{itemize}
Note that we need a sufficient backhaul (not shown in \figref{fig:deployments}) for all deployments except the \indcenBS deployment to permit \nmimo.
Also note that the \glspl{bs} are not necessarily optimally placed.

\subsection{Antenna Array Configuration}
\label{sec:arrayConf}
The indoor \glspl{bs} are rectangle arrays, while the outdoor \glspl{bs} are \glspl{ula}. The antennas are spaced at half wavelength distance $\lambda_L/2$.
The rectangular arrays are mounted underneath the ceiling at a height of \SI{3}{\meter}. We choose the side lengths of the rectangle such that $\lceil \sqrt{M_i} \rceil$ antennas fit per row and column. Note that the last rows might not be fully occupied by antennas.
The height of the outdoor \glspl{bs} is \SI{10}{\meter} and the antennas of the \glspl{ula} are located on a line parallel to the long side of the building.
We assume no mutual coupling between antennas.
Unless otherwise stated, we assume ideal hardware, perfect synchronization, and perfect \gls{csi} of the complete network at all nodes.

\section{Simulation Parameters}
\label{sec:simuPara}
We fix the number of \glspl{ue} to $K=24$ and compare the deployments with different performance measures for different numbers $M$ of total \gls{bs} antennas. We simulate $300$ drops where one drop is a random placement of the $24$ \glspl{ue} within the office building.
For each drop we generate $10$ channel realizations.
The wall penetration loss is \SI{12}{\dB} per wall.
We use a bandwidth of \SI{20}{\MHz} around a carrier frequency of \SI{2.1}{\GHz}. The active bandwidth is \SI{18}{\MHz} and \SI{1}{\MHz} on each side of this bandwidth is a guard band. The subcarrier spacing is \SI{15}{\kHz} and we obtain $1200$ subcarriers.
In \gls{lte}, subcarriers are arranged in groups of $12$ consecutive subcarriers which are called \glspl{prb}. Hence we obtain $100$ \glspl{prb}. The channel conditions of the subcarriers of one \gls{prb} are usually very similar. The schedule, power allocation, and precoder are the same for all subcarriers of one \gls{prb} in \gls{lte} to save control signaling overhead. We save simulation time by simulating a single subcarrier per \gls{prb} and assuming that the same performance is achieved on the other subcarriers of the \gls{prb}.

Unless otherwise mentioned, we use 256 \gls{qam} and \merWf to allocate power. The per-\gls{bs} power in \SI{}{\dBm} at the $i$-th \gls{bs} is constrained by
\begin{equation}
 P_i = \SI{26}{\dBm} - 10 \log_{10}\left(N_{\text{BS}}\right).
\end{equation}
The maximal per-\gls{bs} powers are such that the maximal sum power available to the \glspl{bs} is \SI{26}{\dBm}.
The variance of the \gls{awgn} at the \glspl{ue}, i.e., the noise level, is $\sigma_N^2 = \SI{-125.1}{\dBm}$.

The simulation parameters are summarized in Table \ref{tab:para}.
With these parameters, the per-\gls{ue} \gls{se} $S_k$ of the $k$-th \gls{ue} without considering control signaling overhead is 
\begin{equation}
S_k = \frac{12 \cdot \sum_{f=1}^{100} C\left(\text{SINR}_k^{(f)}\right) \cdot 14 }{ \SI{1}{\ms} \cdot \SI{20}{\MHz}}
\end{equation}
where $12$ is the number of subcarriers per \gls{prb}, $100$ is the number of \glspl{prb}, $14$ is the number of \gls{ofdm} blocks per subframe, \SI{1}{\ms} is the duration of one subframe and $C\left(\text{SINR}_k^{(f)}\right)$ is the capacity at $\text{SINR}_k^{(f)}$ of a memoryless channel with 256 \gls{qam} input and continuous output in bits \cite{Ungerboeck82}. 
The sum \gls{se} $S$ in the building without considering control signaling overhead is
\begin{equation}
 S = \sum_{k=1}^{24} S_k
\end{equation}
where $24$ is the number of \glspl{ue}.
The maximal sum \gls{se} for 256 \gls{qam} is $S^* = \SI{161.28}{\bit\per\second\per\hertz}$, since the rate $C\left(\text{SINR}_k^{(f)}\right)$ is bounded by \SI{8}{\bits} for 256 \gls{qam}.

\begin{table}[t!]
\caption{Simulation Parameters}
\centering
\renewcommand{\arraystretch}{1.2}
\begin{tabular}{|l|c|}
\hline
Carrier frequency & \SI{2.1}{\GHz}\\
Bandwidth & \SI{20}{\MHz}\\
Active bandwidth & \SI{18}{\MHz}\\
Subcarrier spacing & \SI{15}{\kHz}\\
Number of subcarriers & $1200$\\
Number of PRBs & $100$\\
Antenna spacing & $\lambda_L/2$ \\
Indoor wall penetration loss & \SI{12}{\dB}\\
Per-\gls{bs} power constraint $P_i$ & $\SI{26}{\dBm} - 10 \log_{10}\left(N_{\text{BS}}\right)$\\
Noise level $\sigma_N^2$ & \SI{-125.1}{\dBm}\\
Modulation scheme & 256 \gls{qam} \\
Number of \glspl{ue} $K$ & $24$ \\
\hline
Number of drops & 300 \\
Number of channel realizations per drop & 10 \\
\hline
\end{tabular}
\label{tab:para}
\end{table}

\section{Sum Spectral Efficiency}
\label{sec:se}

We first analyze the average sum \gls{se} $S$.
We do not show the \SI{5}{\percent}-tile sum \gls{se} and the \SI{95}{\percent}-tile sum \gls{se} as they follow the same trends.
For the \indcenBS deployment there is only one \gls{bs}, hence the curves for \locPre, \gls{lsmimo} and \nmimo are equal.

Consider the sum \gls{se} achieved with \nmimo (solid curves) in \figref{fig:avgSE256Mer_1}.
The deployments perform poorly for the fully loaded \gls{mimo} system with $M=24$ \gls{bs} antennas. The sum \gls{se} improves significantly when few antennas are added.
Adding more antennas increases the sum \gls{se}, but the gain per additional antenna decreases. A ratio of twice as many \gls{bs} antennas as \glspl{ue} seems to be a good trade-off between achieved sum \gls{se} and number of \gls{bs} antennas.
As expected, the distributed deployments outperform the \indcenBS deployment, except for the \outBS deployment which performs poorly with all transmission schemes.

Next consider the sum \gls{se} achieved with \gls{lsmimo} (dashed curves) in \figref{fig:avgSE256Mer_1}.
Recall that for \gls{lsmimo} at least $M=N_\text{BS}K$ total \gls{bs} antennas are required. Similar to \nmimo, adding more antennas increases the sum \gls{se}, and the gain with each additional antenna decreases. Since \gls{lsmimo} does not require a backhaul one can trade off the costs of a backhaul with the number $M$ of \gls{bs} antennas to achieve the sum \gls{se} of \nmimo with \gls{lsmimo}.

\begin{figure}[t!]
  \centering
%
%
%
%
\begin{tikzpicture}

\begin{axis}[%
width=\plotwidth,
height=\plotheight,
scale only axis,
xmode=log,
xmin=24,
xmax=240,
xtick={24,  48,  96, 192},
xticklabels={24,  48,  96, 192},
xlabel={number $M$ of total \gls{bs} antennas},
xlabel style={inner ysep=\legOffSet,yshift=\legOffSetHalf,},
xmajorgrids,
ymin=25,
ymax=165,
ytick={40, 80, 120, 160},
ylabel={average sum \gls{se} $S$ [bit/s/Hz]},
ymajorgrids,
axis x line*=bottom,
axis y line*=left,
legend style={at={($(current bounding box.south west)!.55!(current bounding box.south east)$)},anchor=north west,draw=black,fill=white,legend cell align=left,font=\small,row sep=-1pt,},
every axis plot/.append style={semithick},
]
\addplot [color=sceOne,solid,mark=\markSceOne,mark options={solid}]
  table[row sep=crcr]{%
24	25.5009436036566\\ 28	44.6897102152222\\ 32	56.1012347553572\\ 36	63.9829172885328\\ 40	70.2558450901997\\ 44	75.2957641845319\\ 48	79.7247505661824\\ 72	96.9115572226586\\ 96	106.81601535717\\ 192	125.053546248337\\ 240	130.247373830157\\ };
\addlegendentry{\legSceOne};

\addplot [color=sceTwo,solid,mark=\markSceTwo,mark options={solid}]
  table[row sep=crcr]{%
24	64.4983133931951\\ 28	112.64838690166\\ 32	128.640654457055\\ 36	136.962574580662\\ 40	142.363456876603\\ 44	146.156511677311\\ 48	148.446426606709\\ 72	156.442652148016\\ 96	159.125334015914\\ 192	161.105383782519\\ 240	161.234799320039\\ };
\addlegendentry{\legSceTwo};

\addplot [color=sceSeven,solid,mark=\markSceSeven,mark options={solid}]
  table[row sep=crcr]{%
24	72.9592816732855\\ 28	132.586006540736\\ 32	147.513145876854\\ 36	153.696157284004\\ 40	157.395361038226\\ 44	158.658984609036\\ 48	159.654472353817\\ 72	161.137920905729\\ 96	161.271004922306\\ 192	161.279958357357\\ 240	161.279999567504\\ };
\addlegendentry{\legSceSeven};

\addplot [color=sceNine,solid,mark=\markSceNine,mark options={solid}]
  table[row sep=crcr]{%
40	144.776169239674\\ 80	161.249436303603\\ 120	161.279999998838\\ 160	161.28000000001\\ 240	161.280000000012\\ };
\addlegendentry{\legSceNine};

\addplot [color=sceTen,solid,mark=\markSceTen,mark options={solid}]
  table[row sep=crcr]{%
24	17.7742566757735\\ 28	38.9500225371682\\ 32	49.325235999242\\ 36	55.9621376034078\\ 40	60.7771067977484\\ 44	64.1918008963444\\ 48	67.8958804802278\\ 72	80.8939598496013\\ 96	88.8890799597981\\ 192	107.368639993274\\ 240	112.081417344803\\ };
\addlegendentry{\legSceTen};

\addplot [color=sceEleven,solid,mark=\markSceEleven,mark options={solid}]
  table[row sep=crcr]{%
24	25.0045785938977\\ 27	47.0581161492597\\ 30	60.8755382352968\\ 33	69.6619238749973\\ 36	75.9075619214897\\ 39	81.5232818945279\\ 42	85.1726203947044\\ 45	88.8274572330861\\ 48	91.8080302236027\\ 72	108.84705556904\\ 96	118.150701227587\\ 192	134.033525446036\\ 240	138.22487363473\\ };
\addlegendentry{\legSceEleven};

%
%

\addplot [color=sceTwo,densely dotted,mark=\markSceTwo,mark options={solid},forget plot]
  table[row sep=crcr]{%
24	88.2004503052445\\ 28	93.7420243773975\\ 32	100.423402631001\\ 36	106.767681362514\\ 40	111.458981808752\\ 44	115.096971013084\\ 48	117.192164698763\\ 72	125.079081450163\\ 96	128.567669172609\\ 192	134.778200710812\\ 240	136.632538269761\\ };

\addplot [color=sceSeven,densely dotted,mark=\markSceSeven,mark options={solid},forget plot]
  table[row sep=crcr]{%
24	56.3800362695434\\ 28	58.9231225335877\\ 32	61.4755036742962\\ 36	64.8825571216359\\ 40	67.4493609434151\\ 44	69.6727522908206\\ 48	71.8542534963476\\ 72	80.8596323370482\\ 96	86.048460475942\\ 192	97.0802033231694\\ 240	101.179062042786\\ };

\addplot [color=sceNine,densely dotted,mark=\markSceNine,mark options={solid},forget plot]
  table[row sep=crcr]{%
40	92.4390548807616\\ 80	110.474404375142\\ 120	119.955064817295\\ 160	125.816253509033\\ 240	133.038946463657\\ };

\addplot [color=sceTen,densely dotted,mark=\markSceTen,mark options={solid},forget plot]
  table[row sep=crcr]{%
24	36.3261965896631\\ 28	41.6940297195658\\ 32	48.9024594888214\\ 36	54.7883501444098\\ 40	60.4003867457908\\ 44	64.2506042614584\\ 48	68.1122814160545\\ 72	81.2008409100235\\ 96	88.9982323256916\\ 192	106.002591046541\\ 240	111.069345921744\\ };

\addplot [color=sceEleven,densely dotted,mark=\markSceEleven,mark options={solid},forget plot]
  table[row sep=crcr]{%
24	46.5840174921098\\ 27	48.997066195877\\ 30	51.6631522111537\\ 33	53.3573124019219\\ 36	55.2594662098772\\ 39	57.7646445044896\\ 42	59.4065556276856\\ 45	60.6633558022401\\ 48	62.3163943966439\\ 72	75.8240321141341\\ 96	84.6860610194119\\ 192	100.291302464506\\ 240	105.268776808814\\ };

%
%
%

\end{axis}

\node[draw, align=left, fill=white, anchor=south east, font=\small] at ($(current bounding box.south west)!.52!(current bounding box.south east)$) {solid: \nmimo \\ dotted: \locPre};

\end{tikzpicture}%
  \caption{Average sum \glspl{se} with \nmimo and \locPre for 256 \gls{qam} and \merWf.}
  \label{fig:avgSE256Mer_2}
\end{figure}
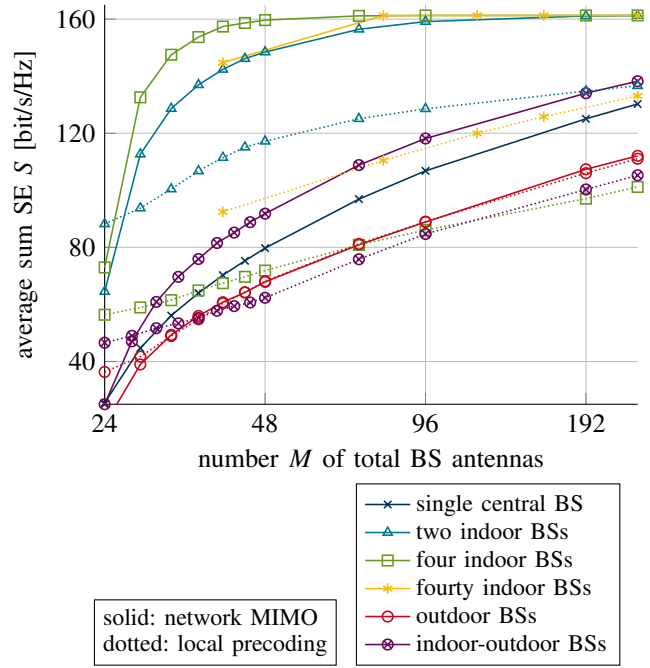

\LocPre is non-cooperative and performs poorly due to interference (dotted curves), see \figref{fig:avgSE256Mer_2}. For all deployments the sum \gls{se} improves little when adding antennas.
However, it may be beneficial to distribute \gls{bs} antennas even without cooperation. For example, the \indtwoBS deployment with \locPre outperforms the \indcenBS deployment. 
\LocPre outperforms \nmimo for small $M$ when more \glspl{ue} are served by a \gls{bs} than the \gls{bs} can serve with \locPre and only the best \glspl{ue} are scheduled.


\begin{figure}[t!]
  \centering
%
%
%
%
\begin{tikzpicture}

\begin{axis}[%
width=\plotwidth,
height=\plotheight,
scale only axis,
xmode=log,
xmin=24,
xmax=240,
xtick={24,  48,  96, 192},
xticklabels={24,  48,  96, 192},
xlabel={number $M$ of total \gls{bs} antennas},
xlabel style={inner ysep=\legOffSet,yshift=\legOffSetHalf,},
xmajorgrids,
ymin=25,
ymax=165,
ytick={40, 80, 120, 160},
ylabel={average sum \gls{se} $S$ [bit/s/Hz]},
ymajorgrids,
axis x line*=bottom,
axis y line*=left,
legend style={at={($(current bounding box.south west)!.55!(current bounding box.south east)$)},anchor=north west,draw=black,fill=white,legend cell align=left,font=\small,row sep=-1pt,},
every axis plot/.append style={semithick},
]
\addplot [color=sceOne,solid,mark=\markSceOne,mark options={solid}]
  table[row sep=crcr]{%
24	25.5009436036566\\ 28	44.6897102152222\\ 32	56.1012347553572\\ 36	63.9829172885328\\ 40	70.2558450901997\\ 44	75.2957641845319\\ 48	79.7247505661824\\ 72	96.9115572226586\\ 96	106.81601535717\\ 192	125.053546248337\\ 240	130.247373830157\\ };
\addlegendentry{\legSceOne};

\addplot [color=sceTwo,solid,mark=\markSceTwo,mark options={solid}]
  table[row sep=crcr]{%
24	64.4983133931951\\ 28	112.64838690166\\ 32	128.640654457055\\ 36	136.962574580662\\ 40	142.363456876603\\ 44	146.156511677311\\ 48	148.446426606709\\ 72	156.442652148016\\ 96	159.125334015914\\ 192	161.105383782519\\ 240	161.234799320039\\ };
\addlegendentry{\legSceTwo};

\addplot [color=sceSeven,solid,mark=\markSceSeven,mark options={solid}]
  table[row sep=crcr]{%
24	72.9592816732855\\ 28	132.586006540736\\ 32	147.513145876854\\ 36	153.696157284004\\ 40	157.395361038226\\ 44	158.658984609036\\ 48	159.654472353817\\ 72	161.137920905729\\ 96	161.271004922306\\ 192	161.279958357357\\ 240	161.279999567504\\ };
\addlegendentry{\legSceSeven};

\addplot [color=sceNine,solid,mark=\markSceNine,mark options={solid}]
  table[row sep=crcr]{%
40	144.776169239674\\ 80	161.249436303603\\ 120	161.279999998838\\ 160	161.28000000001\\ 240	161.280000000012\\ };
\addlegendentry{\legSceNine};

\addplot [color=sceTen,solid,mark=\markSceTen,mark options={solid}]
  table[row sep=crcr]{%
24	17.7742566757735\\ 28	38.9500225371682\\ 32	49.325235999242\\ 36	55.9621376034078\\ 40	60.7771067977484\\ 44	64.1918008963444\\ 48	67.8958804802278\\ 72	80.8939598496013\\ 96	88.8890799597981\\ 192	107.368639993274\\ 240	112.081417344803\\ };
\addlegendentry{\legSceTen};

\addplot [color=sceEleven,solid,mark=\markSceEleven,mark options={solid}]
  table[row sep=crcr]{%
24	25.0045785938977\\ 27	47.0581161492597\\ 30	60.8755382352968\\ 33	69.6619238749973\\ 36	75.9075619214897\\ 39	81.5232818945279\\ 42	85.1726203947044\\ 45	88.8274572330861\\ 48	91.8080302236027\\ 72	108.84705556904\\ 96	118.150701227587\\ 192	134.033525446036\\ 240	138.22487363473\\ };
\addlegendentry{\legSceEleven};

\addplot [color=sceTwo,dashed,mark=\markSceTwo,mark options={solid},forget plot]
  table[row sep=crcr]{%
48	78.9811013889389\\ 50	93.8585625670603\\ 52	104.95085730711\\ 56	116.898778119951\\ 60	125.503671084794\\ 66	134.174974196333\\ 72	140.477134489428\\ 84	148.082703128943\\ 96	152.545550906411\\ 120	157.148028066005\\ 192	160.626713139697\\ 240	161.071409493392\\ };

\addplot [color=sceSeven,dashed,mark=\markSceSeven,mark options={solid},forget plot]
  table[row sep=crcr]{%
96	104.368505677397\\ 100	121.653645363239\\ 104	133.589418711656\\ 108	139.468103683831\\ 112	143.389635401679\\ 120	149.363205192449\\ 128	153.741013921178\\ 140	156.85324540873\\ 160	159.389610698151\\ 192	160.738954421448\\ 240	161.202790956867\\ };

\addplot [color=sceTen,dashed,mark=\markSceTen,mark options={solid},forget plot]
  table[row sep=crcr]{%
48	28.093519274984\\ 50	37.6029846085284\\ 52	43.6039261046155\\ 56	52.5192784885978\\ 60	57.6088056357732\\ 66	64.3464342034445\\ 72	68.921634557646\\ 84	76.9611728591308\\ 96	81.9935048718177\\ 120	90.4464447100418\\ 192	104.60622428149\\ 240	110.546217090343\\ };

\addplot [color=sceEleven,dashed,mark=\markSceEleven,mark options={solid},forget plot]
  table[row sep=crcr]{%
72	41.421842023709\\ 75	53.1134335695491\\ 78	61.6746508112146\\ 84	72.2406368870643\\ 90	80.5818914682623\\ 96	86.498246275265\\ 108	94.5492377042407\\ 120	101.952445829009\\ 141	110.652212626186\\ 192	122.961812667059\\ 240	129.590578981422\\ };

\end{axis}

\node[draw, align=left, fill=white, anchor=south east, font=\small] at ($(current bounding box.south west)!.52!(current bounding box.south east)$) {solid: \nmimo \\ dashed: \gls{lsmimo}};

\end{tikzpicture}%
  \caption{Average sum \glspl{se} with \nmimo and \gls{lsmimo} for 256 \gls{qam} and \merWf.}
  \label{fig:avgSE256Mer_1}
\end{figure}

In conclusion, the \gls{se} increases with the number of \gls{bs} antennas for all deployments and all transmission schemes until it is limited by the maximal \gls{se} of the modulation. Cooperation between indoor \glspl{bs} provides large gains, while cooperation between outdoor \glspl{bs}, or indoor and outdoor \glspl{bs} provides smaller gains. \Nmimo performs best, but \gls{cscb} is an interesting alternative as the backhaul requirements are reduced. The placement of \glspl{bs} is important to overcome wall penetration losses and to control interference.

\section{Average SNR Maps}
\label{sec:SNRmrt}
In this section, we show why the deployments with only one or no indoor \gls{bs} perform poorly as compared to the distributed indoor \glspl{bs} deployments.
We analyze the \gls{snr} achieved when only a single \gls{ue} is served at different positions within the office building. The \glspl{bs} use \nmimo under per-\gls{bs} power constraints.\footnote{For a single served \gls{ue}, \gls{zfbf} coincides with maximum ratio transmission.} We distribute the per-\gls{bs} transmit power equally among the subcarriers.
The \gls{snr} achieved when a single \gls{ue} is served is an upper bound to the \gls{snr} when more \glspl{ue} are served with \gls{zfbf} or any other linear precoding scheme, as serving more \glspl{ue} only reduces the degrees-of-freedom.

\begin{figure*}[t!]
  \centering
  \subfloat[\hspace*{3em} (a) \IndcenBS deployment.]{
%
%
\begin{tikzpicture}[font=\small]

\begin{axis}[%
width=\halfmapwidth,
height=\halfmapheight,
scale only axis,
point meta min=5,
point meta max=95,
xmin=-50,
xmax=50,
xtick={-40, 0,  40},
xticklabels={$\SI{-40}{\meter}$, $\SI{0}{\meter}$, $\SI{40}{\meter}$},
ymin=-25,
ymax=25,
ytick={-20, 0, 20},
yticklabels={$\SI{-20}{\meter}$, $\SI{0}{\meter}$, $\SI{20}{\meter}$},
axis background/.style={fill=white},
axis line style = very thin,
axis x line*=bottom,
axis y line*=left,
]
\addplot [forget plot] graphics [xmin=-50,xmax=50,ymin=-25,ymax=25] {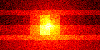};
\addplot [draw=none,fill=black,forget plot]
  table[row sep=crcr]{%
0	0\\
0	-2.5\\
-2.5	-2.5\\
-2.5	0\\
0	0\\
};

\input{./fig/twoStripe.tikz}
\end{axis}
\end{tikzpicture}%
 \label{fig:mrtsnr1}}
  \hfill
  \subfloat[(b) \IndtwoBS deployment.]{
%
%
\begin{tikzpicture}[font=\small]

\begin{axis}[%
width=\halfmapwidth,
height=\halfmapheight,
scale only axis,
point meta min=5,
point meta max=95,
xmin=-50,
xmax=50,
xtick={-40, 0,  40},
xticklabels={$\SI{-40}{\meter}$, $\SI{0}{\meter}$, $\SI{40}{\meter}$},
ymin=-25,
ymax=25,
ytick={-20, 0, 20},
yticklabels={,,},
axis background/.style={fill=white},
axis line style = very thin,
axis x line*=bottom,
axis y line*=left,
]
\addplot [forget plot] graphics [xmin=-50,xmax=50,ymin=-25,ymax=25] {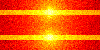};
\addplot [draw=none,fill=black,forget plot]
  table[row sep=crcr]{%
-.9	11.6\\
-.9	13.4\\
.9	13.4\\
.9	11.6\\
-.9	11.6\\
};
\addplot [draw=none,fill=black,forget plot]
  table[row sep=crcr]{%
-.9	-11.6\\
-.9	-13.4\\
.9	-13.4\\
.9	-11.6\\
-.9	-11.6\\
};

\input{./fig/twoStripe.tikz}
\end{axis}
\end{tikzpicture}%
 \label{fig:mrtsnr2}}
  \hfill
  \subfloat[(c) \IndfourBS deployment. \hspace*{6em}]{
%
%
\begin{tikzpicture}[font=\small]

\begin{axis}[%
width=\halfmapwidth,
height=\halfmapheight,
scale only axis,
point meta min=5,
point meta max=95,
xmin=-50,
xmax=50,
xtick={-40,  0,   40},
xticklabels={$\SI{-40}{\meter}$, $\SI{0}{\meter}$, $\SI{40}{\meter}$},
ymin=-25,
ymax=25,
ytick={-20, 0, 20},
yticklabels={,,},
axis background/.style={fill=white},
axis line style = very thin,
axis x line*=bottom,
axis y line*=left,
colormap/hot2,
colorbar,
colorbar style = {ylabel=SNR [\SI{}{\dB}],ytick={10,30,50,70,90},},
]
\addplot [forget plot] graphics [xmin=-50,xmax=50,ymin=-25,ymax=25] {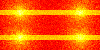};
\addplot [draw=none,fill=black,forget plot]
  table[row sep=crcr]{%
-30.63	11.87\\
-30.63	13.13\\
-29.37	13.13\\
-29.37	11.87\\
-30.63	11.87\\
};
\addplot [draw=none,fill=black,forget plot]
  table[row sep=crcr]{%
30.63	-11.87\\
30.63	-13.13\\
29.37	-13.13\\
29.37	-11.87\\
30.63	-11.87\\
};
\addplot [draw=none,fill=black,forget plot]
  table[row sep=crcr]{%
-30.63	-11.87\\
-30.63	-13.13\\
-29.37	-13.13\\
-29.37	-11.87\\
-30.63	-11.87\\
};
\addplot [draw=none,fill=black,forget plot]
  table[row sep=crcr]{%
30.63	11.87\\
30.63	13.13\\
29.37	13.13\\
29.37	11.87\\
30.63	11.87\\
};

\input{./fig/twoStripe.tikz}
\end{axis}
\end{tikzpicture}%
 \label{fig:mrtsnr7}}
  \\
  \subfloat[\hspace*{3em} (d) \IndfourtyBS deployment.]{
%
%
\begin{tikzpicture}[font=\small]

\begin{axis}[%
width=\halfmapwidth,
height=\halfmapheight,
scale only axis,
point meta min=5,
point meta max=95,
xmin=-50,
xmax=50,
xtick={-40, 0,  40},
xticklabels={$\SI{-40}{\meter}$, $\SI{0}{\meter}$, $\SI{40}{\meter}$},
ymin=-25,
ymax=25,
ytick={-20, 0, 20},
yticklabels={$\SI{-20}{\meter}$, $\SI{0}{\meter}$, $\SI{20}{\meter}$},
axis background/.style={fill=white},
axis line style = very thin,
axis x line*=bottom,
axis y line*=left,
]
\addplot [forget plot] graphics [xmin=-50,xmax=50,ymin=-25,ymax=25] {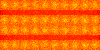};

\filldraw[fill=black, draw=black] (axis cs: -45.2,-20.2) rectangle (axis cs: -44.8,-19.8);
\filldraw[fill=black, draw=black] (axis cs: -35.2,-20.2) rectangle (axis cs: -34.8,-19.8);
\filldraw[fill=black, draw=black] (axis cs: -25.2,-20.2) rectangle (axis cs: -24.8,-19.8);
\filldraw[fill=black, draw=black] (axis cs: -15.2,-20.2) rectangle (axis cs: -14.8,-19.8);
\filldraw[fill=black, draw=black] (axis cs: -5.2,-20.2) rectangle (axis cs: -4.8,-19.8);
\filldraw[fill=black, draw=black] (axis cs: 4.8,-20.2) rectangle (axis cs: 5.2,-19.8);
\filldraw[fill=black, draw=black] (axis cs: 14.8,-20.2) rectangle (axis cs: 15.2,-19.8);
\filldraw[fill=black, draw=black] (axis cs: 24.8,-20.2) rectangle (axis cs: 25.2,-19.8);
\filldraw[fill=black, draw=black] (axis cs: 34.8,-20.2) rectangle (axis cs: 35.2,-19.8);
\filldraw[fill=black, draw=black] (axis cs: 44.8,-20.2) rectangle (axis cs: 45.2,-19.8);
 
\filldraw[fill=black, draw=black] (axis cs: -45.2,-5.2) rectangle (axis cs: -44.8,-4.8);
\filldraw[fill=black, draw=black] (axis cs: -35.2,-5.2) rectangle (axis cs: -34.8,-4.8);
\filldraw[fill=black, draw=black] (axis cs: -25.2,-5.2) rectangle (axis cs: -24.8,-4.8);
\filldraw[fill=black, draw=black] (axis cs: -15.2,-5.2) rectangle (axis cs: -14.8,-4.8);
\filldraw[fill=black, draw=black] (axis cs: -5.2,-5.2) rectangle (axis cs: -4.8,-4.8);
\filldraw[fill=black, draw=black] (axis cs: 4.8,-5.2) rectangle (axis cs: 5.2,-4.8);
\filldraw[fill=black, draw=black] (axis cs: 14.8,-5.2) rectangle (axis cs: 15.2,-4.8);
\filldraw[fill=black, draw=black] (axis cs: 24.8,-5.2) rectangle (axis cs: 25.2,-4.8);
\filldraw[fill=black, draw=black] (axis cs: 34.8,-5.2) rectangle (axis cs: 35.2,-4.8);
\filldraw[fill=black, draw=black] (axis cs: 44.8,-5.2) rectangle (axis cs: 45.2,-4.8);

\filldraw[fill=black, draw=black] (axis cs: -45.2,5.2) rectangle (axis cs: -44.8,4.8);
\filldraw[fill=black, draw=black] (axis cs: -35.2,5.2) rectangle (axis cs: -34.8,4.8);
\filldraw[fill=black, draw=black] (axis cs: -25.2,5.2) rectangle (axis cs: -24.8,4.8);
\filldraw[fill=black, draw=black] (axis cs: -15.2,5.2) rectangle (axis cs: -14.8,4.8);
\filldraw[fill=black, draw=black] (axis cs: -5.2,5.2) rectangle (axis cs: -4.8,4.8);
\filldraw[fill=black, draw=black] (axis cs: 4.8,5.2) rectangle (axis cs: 5.2,4.8);
\filldraw[fill=black, draw=black] (axis cs: 14.8,5.2) rectangle (axis cs: 15.2,4.8);
\filldraw[fill=black, draw=black] (axis cs: 24.8,5.2) rectangle (axis cs: 25.2,4.8);
\filldraw[fill=black, draw=black] (axis cs: 34.8,5.2) rectangle (axis cs: 35.2,4.8);
\filldraw[fill=black, draw=black] (axis cs: 44.8,5.2) rectangle (axis cs: 45.2,4.8);

\filldraw[fill=black, draw=black] (axis cs: -45.2,20.2) rectangle (axis cs: -44.8,19.8);
\filldraw[fill=black, draw=black] (axis cs: -35.2,20.2) rectangle (axis cs: -34.8,19.8);
\filldraw[fill=black, draw=black] (axis cs: -25.2,20.2) rectangle (axis cs: -24.8,19.8);
\filldraw[fill=black, draw=black] (axis cs: -15.2,20.2) rectangle (axis cs: -14.8,19.8);
\filldraw[fill=black, draw=black] (axis cs: -5.2,20.2) rectangle (axis cs: -4.8,19.8);
\filldraw[fill=black, draw=black] (axis cs: 4.8,20.2) rectangle (axis cs: 5.2,19.8);
\filldraw[fill=black, draw=black] (axis cs: 14.8,20.2) rectangle (axis cs: 15.2,19.8);
\filldraw[fill=black, draw=black] (axis cs: 24.8,20.2) rectangle (axis cs: 25.2,19.8);
\filldraw[fill=black, draw=black] (axis cs: 34.8,20.2) rectangle (axis cs: 35.2,19.8);
\filldraw[fill=black, draw=black] (axis cs: 44.8,20.2) rectangle (axis cs: 45.2,19.8);

\input{./fig/twoStripe.tikz}

\end{axis}
\end{tikzpicture}%
 \label{fig:mrtsnr9}}
  \hfill
  \subfloat[(e) \OutBS deployment.]{
%
%
\begin{tikzpicture}[font=\small]

\begin{axis}[%
width=\halfmapwidth,
height=\halfmapheight,
scale only axis,
point meta min=5,
point meta max=95,
xmin=-50,
xmax=50,
xtick={-40, 0,  40},
xticklabels={$\SI{-40}{\meter}$, $\SI{0}{\meter}$, $\SI{40}{\meter}$},
ymin=-25,
ymax=25,
ytick={-20, 0, 20},
yticklabels={,,,,},
axis background/.style={fill=white},
axis line style = very thin,
axis x line*=bottom,
axis y line*=left,
]
\addplot [forget plot] graphics [xmin=-50,xmax=50,ymin=-25,ymax=25] {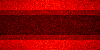};

\addplot [draw=none,fill=black,forget plot]
  table[row sep=crcr]{%
3.2	40\\
3.2	39.5\\
-3.2	39.5\\
-3.2	40\\
3.2	40\\
};
\addplot [draw=none,fill=black,forget plot]
  table[row sep=crcr]{%
3.2	-40\\
3.2	-39.5\\
-3.2	-39.5\\
-3.2	-40\\
3.2	-40\\
};

\input{./fig/twoStripe.tikz}
\end{axis}
\end{tikzpicture}%
 \label{fig:mrtsnr10}}
  \hfill
  \subfloat[(f) \InoutBS deployments.\hspace*{6em}]{
%
%
\begin{tikzpicture}[font=\small]

\begin{axis}[%
width=\halfmapwidth,
height=\halfmapheight,
scale only axis,
point meta min=5,
point meta max=95,
xmin=-50,
xmax=50,
xtick={-40, 0,  40},
xticklabels={$\SI{-40}{\meter}$, $\SI{0}{\meter}$, $\SI{40}{\meter}$},
ymin=-25,
ymax=25,
ytick={-20, 0, 20},
yticklabels={,,,,},
axis background/.style={fill=white},
axis line style = very thin,
axis x line*=bottom,
axis y line*=left,
colormap/hot2,
colorbar,
colorbar style = {ylabel=SNR [\SI{}{\dB}],ytick={10,30,50,70,90},},
]
\addplot [forget plot] graphics [xmin=-50,xmax=50,ymin=-25,ymax=25] {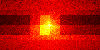};
\addplot [draw=none,fill=black,forget plot]
  table[row sep=crcr]{%
2.14	40\\
2.14	39.5\\
-2.14	39.5\\
-2.14	40\\
2.14	40\\
};
\addplot [draw=none,fill=black,forget plot]
  table[row sep=crcr]{%
2.14	-40\\
2.14	-39.5\\
-2.14	-39.5\\
-2.14	-40\\
2.14	-40\\
};
\addplot [draw=none,fill=black,forget plot]
  table[row sep=crcr]{%
0	0\\
0	-1.44\\
-1.44	-1.44\\
-1.44	0\\
0	0\\
};

\input{./fig/twoStripe.tikz}
\end{axis}
\end{tikzpicture}%
 \label{fig:mrtsnr11}}
  \caption{Average SNR achieved at a single served \gls{ue} for different positions with 48 transmit antennas (with 40 transmit antennas for the \indfourtyBS deployment).}
  \label{fig:mrtsnr}
\end{figure*}

\figref{fig:mrtsnr} shows the \glspl{snr} averaged over $300$ channel realizations for each sampled position.
The \indcenBS deployment achieves low \gls{snr} in many rooms, especially those close to the outside wall. This is due to the wall penetration loss. 
The \outBS deployment and the \inoutBS deployment achieve low \gls{snr} in inner rooms and in the corridors.
The other deployments achieve high \gls{snr} in all rooms.
We conclude that the lower \glspl{se} of the deployments with only one or no indoor \gls{bs} are at least partly due to the large wall penetration loss and the building penetration loss. A deployment with few well-placed \glspl{bs} suffices to provide good service throughout the building.

\section{Comparison to Capacity Upper Bound}
\label{sec:capSim}
\Mmimo lets simple transmission schemes approach capacity with an increasing number of \gls{bs} antennas. In the following, we analyze this statement for the \nmimo transmission scheme.
We upper bound the capacity of a deployment by the capacity of a \gls{bc} under a total power constraint. 
We allow all \glspl{bs} of a deployment to cooperate and to act as one \gls{bs} with distributed antennas, and relax the per-\gls{bs} power constraint to a total-power constraint.
Note that for the \indcenBS deployment the upper bound is tight, as the capacity of a \gls{bc} is achieved by non-linear \gls{dpc} \cite{Caire03,ViswanathTse03,VishwJind03,Yu04}.
We find the optimal transmission policy with the algorithms in \cite{jindal05} treating the \gls{ofdm} subcarriers as virtual antennas.
We compare capacity to the \glspl{se} achieved with Gaussian modulation, since 256 \gls{qam} limits \gls{se}, while Gaussian modulation allows to approach the capacity upper bound.

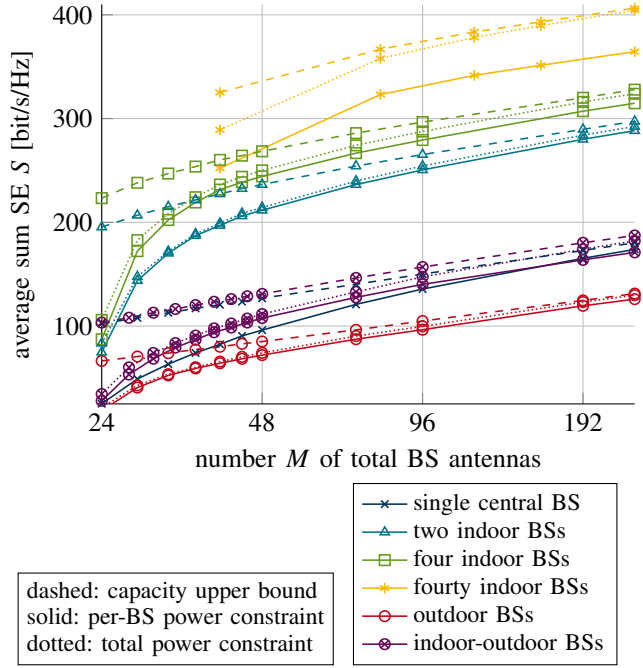
\begin{figure}[t!]
  \centering
%
%
%
%
\begin{tikzpicture}

\begin{axis}[%
width=\plotwidth,
height=\plotheight,
scale only axis,
xmode=log,
xmin=24,
xmax=240,
xtick={24, 48, 96, 192},
xticklabels={24, 48, 96, 192},
xlabel={number $M$ of total \gls{bs} antennas},
xlabel style={inner ysep=\legOffSet,yshift=\legOffSetHalf,},
xmajorgrids,
ymin=25,
ymax=410,
ytick={100, 200, 300, 400},
ylabel={average sum \gls{se} $S$ [bit/s/Hz]},
ymajorgrids,
axis x line*=bottom,
axis y line*=left,
legend style={at={($(current bounding box.south west)!.55!(current bounding box.south east)$)},anchor=north west,draw=black,fill=white,legend cell align=left,font=\small,row sep=-1pt,},
every axis plot/.append style={semithick},
]
\addplot [color=sceOne,solid,mark=\markSceOne,mark options={solid}]
  table[row sep=crcr]{%
24	25.1938991546732\\ 28	48.9941607939292\\ 32	63.3463545817879\\ 36	74.1375285196521\\ 40	82.4095751118175\\ 44	90.4912538787455\\ 48	96.1298245692978\\ 72	120.901233236184\\ 96	135.696134818841\\ 192	165.326475060043\\ 240	173.882531877858\\ };
\addlegendentry{\legSceOne};

\addplot [color=sceTwo,solid,mark=\markSceTwo,mark options={solid}]
  table[row sep=crcr]{%
24	74.8261186341351\\ 28	143.805783229715\\ 32	170.219670992785\\ 36	187.23550986486\\ 40	196.968461281446\\ 44	206.163105590242\\ 48	211.659522433162\\ 72	236.378331209789\\ 96	250.535777432824\\ 192	280.032490922534\\ 240	288.343492338342\\ };
\addlegendentry{\legSceTwo};

\addplot [color=sceSeven,solid,mark=\markSceSeven,mark options={solid}]
  table[row sep=crcr]{%
24	87.2607076701511\\ 28	172.241156273197\\ 32	202.245519810037\\ 36	219.141404329109\\ 40	231.528058712644\\ 44	238.674739393692\\ 48	244.286903610131\\ 72	266.71093852066\\ 96	279.355032351265\\ 192	307.137350171565\\ 240	314.766046145203\\ };
\addlegendentry{\legSceSeven};

\addplot [color=sceNine,solid,mark=\markSceNine,mark options={solid}]
  table[row sep=crcr]{%
40	252.181174366846\\ 80	323.204322677917\\ 120	341.67487617444\\ 160	351.423961345268\\ 240	364.535198882565\\ };
\addlegendentry{\legSceNine};

\addplot [color=sceTen,solid,mark=\markSceTen,mark options={solid}]
  table[row sep=crcr]{%
  24	18.4444385975474\\ 28	40.7139352924878\\ 32	52.4675709851002\\ 36	59.1858682241224\\ 40	64.1956508044593\\ 44	68.4741143069837\\ 48	72.0183947520367\\ 72	87.3661673728205\\ 96	96.5102812508564\\ 192	119.540019167341\\ 240	125.972792193835\\ };
\addlegendentry{\legSceTen};

\addplot [color=sceEleven,solid,mark=\markSceEleven,mark options={solid}]
  table[row sep=crcr]{%
  24	27.960819465918\\ 27	53.3240424864021\\ 30	68.5469229472445\\ 33	79.3009312102554\\ 36	86.8461956607725\\ 39	94.0322368695246\\ 42	98.5905282008209\\ 45	103.451535519142\\ 48	107.767740586839\\ 72	127.642837940773\\ 96	140.241911517048\\ 192	163.79840222651\\ 240	171.050436887436\\ };
 \addlegendentry{\legSceEleven};

\addplot [color=sceTwo,densely dotted,mark=\markSceTwo,mark options={solid}]
  table[row sep=crcr]{%
24	83.4219564141106\\ 28	147.648586730221\\ 32	172.010940786919\\ 36	188.935602925994\\ 40	199.086168388347\\ 44	208.662065435404\\ 48	214.341095412297\\ 72	239.870277544504\\ 96	254.312396333089\\ 192	284.053874872178\\ 240	292.402109305614\\ };

\addplot [color=sceSeven,densely dotted,mark=\markSceSeven,mark options={solid}]
  table[row sep=crcr]{%
24	105.960382700742\\ 28	182.819224921539\\ 32	208.459125313889\\ 36	224.196723496846\\ 40	236.419606874622\\ 44	243.918993575994\\ 48	250.067607316015\\ 72	274.216153834223\\ 96	287.567391343358\\ 192	316.097438468627\\ 240	323.925108460632\\ };

\addplot [color=sceNine,densely dotted,mark=\markSceNine,mark options={solid}]
  table[row sep=crcr]{%
40	289.154552286934\\ 80	357.92269082614\\ 120	378.206991644826\\ 160	389.681411187574\\ 240	404.343875751481\\ };

\addplot [color=sceTen,densely dotted,mark=\markSceTen,mark options={solid}]
  table[row sep=crcr]{%
24	21.339726181969\\ 28	42.8486425543958\\ 32	53.6414790091737\\ 36	60.3269612827786\\ 40	65.7932588767401\\ 44	70.3470057701336\\ 48	74.1743484708992\\ 72	90.3328036658784\\ 96	99.7171301595711\\ 192	123.166138511658\\ 240	129.751717000136\\ };

\addplot [color=sceEleven,densely dotted,mark=\markSceEleven,mark options={solid}]
  table[row sep=crcr]{%
  24	34.5411671890719\\ 27	60.2067044270326\\ 30	74.0812835700077\\ 33	83.5607557024631\\ 36	90.7934168676255\\ 39	97.7442100562599\\ 42	102.463846627852\\ 45	107.239785767413\\ 48	111.589800115738\\ 72	132.943507184604\\ 96	147.458569662685\\ 192	174.145884044673\\ 240	182.086912322591\\ };

\addplot [color=sceOne,dashed,mark=\markSceOne,mark options={solid}]
  table[row sep=crcr]{%
24	102.416441405817\\ 28	107.855890452042\\ 32	112.826946582601\\ 36	117.091274875766\\ 40	120.566693117899\\ 44	123.541179877006\\ 48	126.85754991026\\ 72	140.309033631375\\ 96	150.132548543508\\ 192	172.849714777671\\ 240	180.296615047525\\ };

\addplot [color=sceTwo,dashed,mark=\markSceTwo,mark options={solid}]
  table[row sep=crcr]{%
24	195.445352914693\\ 28	206.82448589317\\ 32	215.03021331126\\ 36	221.650559836229\\ 40	227.411147261351\\ 44	232.525094025358\\ 48	236.478040542081\\ 72	254.236292997721\\ 96	265.340643137258\\ 192	289.649582999783\\ 240	297.315078331492\\ };

\addplot [color=sceSeven,dashed,mark=\markSceSeven,mark options={solid}]
  table[row sep=crcr]{%
24	223.296168759377\\ 28	238.13666310891\\ 32	247.024774632989\\ 36	253.749703503926\\ 40	260.187437369666\\ 44	264.281501136925\\ 48	268.501858561865\\ 72	285.992901959342\\ 96	296.639420452461\\ 192	320.086403813817\\ 240	327.90389190415\\ };

\addplot [color=sceNine,dashed,mark=\markSceNine,mark options={solid}]
  table[row sep=crcr]{%
40	325.194724922624\\ 80	366.887351581621\\ 120	383.508009751325\\ 160	393.256062566458\\ 240	406.481339187264\\ };

\addplot [color=sceTen,dashed,mark=\markSceTen,mark options={solid}]
  table[row sep=crcr]{%
24	66.4367845182193\\ 28	70.6483784833749\\ 32	73.9001864653015\\ 36	77.4448720437998\\ 40	80.0994862859255\\ 44	82.8951179472928\\ 48	85.1024828182194\\ 72	96.3754757617052\\ 96	104.886517959108\\ 192	124.751559072429\\ 240	131.312883818084\\ };

\addplot [color=sceEleven,dashed,mark=\markSceEleven,mark options={solid}]
  table[row sep=crcr]{%
  24	103.377330760085\\ 27	108.055066242844\\ 30	112.896101899759\\ 33	116.466450503874\\ 36	120.231100214167\\ 39	123.42177474493\\ 42	126.0019085783\\ 45	128.735904927915\\ 48	130.990359309765\\ 72	146.40193965125\\ 96	156.912767257309\\ 192	180.308838402909\\ 240	187.305994782062\\ };

\end{axis}

\node[draw, align=left, fill=white, anchor=south east, font=\small] at ($(current bounding box.south west)!.52!(current bounding box.south east)$) {dashed: capacity upper bound \\ solid: per-\gls{bs} power constraint \\ dotted: total power constraint };

\end{tikzpicture}%
  \caption{Average sum \glspl{se} of \nmimo for Gaussian modulation.}
  \label{fig:avgSEcapMMimo}
\end{figure}

\figref{fig:avgSEcapMMimo} shows the capacity upper bounds, and the average sum \glspl{se} achieved with Gaussian modulation and \nmimo under per-\gls{bs} power constraints and under a total power constraint. The general trends are similar to \figref{fig:avgSE256Mer_1} and \figref{fig:avgSE256Mer_2}, but the \glspl{se} increase without bound with the number of \gls{bs} antennas. For few \gls{bs} antennas, the gap between the capacity upper bound and \nmimo is large, but the gap could be reduced by more advanced scheduling.
The channels harden for more \gls{bs} antennas: It becomes optimal to schedule all \glspl{ue} on each subcarrier \cite{Bjoernson16}, and advanced scheduling strategies provide diminishing gains \cite{eusipco}.
With an increasing number of \gls{bs} antennas, the gap decreases and vanishes completely under a total power constraint, while a gap remains under per-\gls{bs} power constraints.
Determining better capacity upper bounds, and choosing better precoding and power allocation under per-\gls{bs} power constraints would reduce the gap. 
In summary, \mmimo allows simple transmission schemes to approach capacity with an increasing number of \gls{bs} antennas in our scenarios.

\section{Fairness Analysis}
\label{sec:fairness}
Our deployments and transmission schemes should provide a fair service to all \glspl{ue} as the channels harden.
We measure fairness quantitatively with Jain's index \cite{jainsindex}
\begin{equation}
J \left(S_1, S_2, \dots, S_K \right) = \frac{\left( \sum_{k=1}^K S_k \right)^2}{K \cdot \sum_{k=1}^K {S_k}^2}.
\end{equation}
Jain's index is $1$ when all \glspl{ue} achieve the same per-\gls{ue} \gls{se} and is $1/K$ when only one \gls{ue} achieves a positive per-\gls{ue} \gls{se}.

\begin{figure}[t!]
  \centering
%
%
%
%
\begin{tikzpicture}

\begin{axis}[%
width=\plotwidth,
height=\plotheight,
scale only axis,
xmode=log,
xmin=24,
xmax=240,
xtick={24,  48,  96, 192},
xticklabels={24,  48,  96, 192},
xlabel={number $M$ of total \gls{bs} antennas},
xlabel style={inner ysep=\legOffSet,yshift=\legOffSetHalf,},
xmajorgrids,
ymin=0.4,
ymax=1,
ytick={0.4, 0.6, 0.8, 1},
ylabel={Jain's fairness index $J$},
ymajorgrids,
axis x line*=bottom,
axis y line*=left,
legend style={at={($(current bounding box.south west)!.55!(current bounding box.south east)$)},anchor=north west,draw=black,fill=white,legend cell align=left,font=\small,row sep=-1pt,},
every axis plot/.append style={semithick},
]
\addplot [color=sceOne,solid,mark=\markSceOne,mark options={solid}]
  table[row sep=crcr]{%
24	0.238636697174683\\ 28	0.391477842706965\\ 32	0.483660383333987\\ 36	0.544152962942662\\ 40	0.590431334250629\\ 44	0.627162426716338\\ 48	0.656722295118817\\ 72	0.760832649309015\\ 96	0.814182356914676\\ 192	0.898996484806644\\ 240	0.920527610047755\\ };
\addlegendentry{\legSceOne};

\addplot [color=sceTwo,solid,mark=\markSceTwo,mark options={solid}]
  table[row sep=crcr]{%
24	0.57042815753462\\ 28	0.845276912750747\\ 32	0.916500570634792\\ 36	0.949263329733858\\ 40	0.966516013756192\\ 44	0.976750801604784\\ 48	0.982066543012622\\ 72	0.996186386037716\\ 96	0.998934656923335\\ 192	0.999978421689084\\ 240	0.999997465167487\\ };
\addlegendentry{\legSceTwo};

\addplot [color=sceSeven,solid,mark=\markSceSeven,mark options={solid}]
  table[row sep=crcr]{%
24	0.669055349978763\\ 28	0.940635346780594\\ 32	0.981673221753028\\ 36	0.993299815712913\\ 40	0.997734335314353\\ 44	0.998862511151811\\ 48	0.999428902032931\\ 72	0.999980382097789\\ 96	0.999999914539677\\ 192	0.999999999974699\\ 240	1.00000000000004\\ };
\addlegendentry{\legSceSeven};

\addplot [color=sceNine,solid,mark=\markSceNine,mark options={solid}]
  table[row sep=crcr]{%
40	0.985541973186993\\ 80	0.999991577995984\\ 120	1\\ 160	0.999999999999999\\ 240	1\\ };
\addlegendentry{\legSceNine};

\addplot [color=sceTen,solid,mark=\markSceTen,mark options={solid}]
  table[row sep=crcr]{%
  24	0.312122700533021\\ 28	0.454090671964728\\ 32	0.517675543557849\\ 36	0.560451683699627\\ 40	0.593085544300844\\ 44	0.614437820765046\\ 48	0.637319882737815\\ 72	0.720896139219568\\ 96	0.770420632992164\\ 192	0.860202043293229\\ 240	0.881591957497334\\ };
\addlegendentry{\legSceTen};

\addplot [color=sceEleven,solid,mark=\markSceEleven,mark options={solid}]
  table[row sep=crcr]{%
  24	0.334708700259877\\ 27	0.513435528041108\\ 30	0.605891004879374\\ 33	0.660251886195624\\ 36	0.692836189481242\\ 39	0.722856964458493\\ 42	0.742015238596076\\ 45	0.760434275183852\\ 48	0.778119793077295\\ 72	0.853613462350641\\ 96	0.890187434558949\\ 192	0.94578663461739\\ 240	0.958804672535872\\ };
\addlegendentry{\legSceEleven};

\addplot [color=sceTwo,dashed,mark=\markSceTwo,mark options={solid},forget plot]
  table[row sep=crcr]{%
24	nan\\ 28	nan\\ 32	nan\\ 36	nan\\ 40	nan\\ 44	nan\\ 48	0.643663591022969\\ 50	0.734392873020206\\ 52	0.799675689214698\\ 56	0.864280693085584\\ 60	0.90302513679835\\ 66	0.938093831217049\\ 72	0.958241043828726\\ 84	0.978812411443537\\ 96	0.988479842813075\\ 120	0.996153166444869\\ 192	0.999736666214975\\ 240	0.999952945728066\\ }; 

\addplot [color=sceSeven,dashed,mark=\markSceSeven,mark options={solid},forget plot]
  table[row sep=crcr]{%
24	nan\\ 28	nan\\ 32	nan\\ 36	nan\\ 40	nan\\ 44	nan\\ 48	nan\\ 72	nan\\ 96	0.823513083859689\\ 100	0.893000491899879\\ 104	0.936350403103886\\ 108	0.955302964894946\\ 112	0.965980244211582\\ 120	0.980942051569417\\ 128	0.990009653743144\\ 140	0.99499686252992\\ 160	0.998475682959038\\ 192	0.999736680627946\\ 240	0.999984162802631\\ };

\addplot [color=sceTen,dashed,mark=\markSceTen,mark options={solid},forget plot]
  table[row sep=crcr]{%
  48	0.415886892462441\\ 50	0.459447836021569\\ 52	0.488698440647758\\ 56	0.533591159735975\\ 60	0.567914023123433\\ 66	0.601073709364037\\ 72	0.633279407993681\\ 84	0.678717781645518\\ 96	0.709121942282389\\ 120	0.759535850418685\\ 192	0.833920599279886\\ 240	0.859889194591162\\ };

\addplot [color=sceEleven,dashed,mark=\markSceEleven,mark options={solid},forget plot]
  table[row sep=crcr]{%
  72	0.466300622293044\\ 75	0.534163314910529\\ 78	0.58259466351694\\ 84	0.636063762899724\\ 90	0.679210126911761\\ 96	0.709617428783093\\ 108	0.747171365847173\\ 120	0.787047302928192\\ 141	0.825577977602037\\ 192	0.881952355884363\\ 240	0.909442628437332\\ };

\addplot [color=sceTwo,densely dotted,mark=\markSceTwo,mark options={solid},forget plot]
  table[row sep=crcr]{%
24	0.706409246332825\\ 28	0.742216231034818\\ 32	0.778141987331173\\ 36	0.815295484619217\\ 40	0.839454980234136\\ 44	0.857141176678475\\ 48	0.866571349845338\\ 72	0.893669824708507\\ 96	0.904701476890975\\ 192	0.924801607915467\\ 240	0.931976155523674\\ }; 

\addplot [color=sceSeven,densely dotted,mark=\markSceSeven,mark options={solid},forget plot]
  table[row sep=crcr]{%
24	0.577571363548631\\ 28	0.602303391379069\\ 32	0.61712986892576\\ 36	0.636339498835497\\ 40	0.652755983480427\\ 44	0.670888818881585\\ 48	0.683103752142682\\ 72	0.728222020955579\\ 96	0.748512347916756\\ 192	0.788074481427119\\ 240	0.801031510853492\\ };

\addplot [color=sceNine,densely dotted,mark=\markSceNine,mark options={solid},forget plot]
  table[row sep=crcr]{%
40	0.699573534422744\\ 80	0.815062699431096\\ 120	0.862760810289351\\ 160	0.889543483775916\\ 240	0.919893241293241\\ };

\addplot [color=sceTen,densely dotted,mark=\markSceTen,mark options={solid},forget plot]
  table[row sep=crcr]{%
24	0.436332047594415\\ 28	0.465346200920728\\ 32	0.507973443424435\\ 36	0.542783925431009\\ 40	0.57441093488094\\ 44	0.598995979188711\\ 48	0.620807088389176\\ 72	0.698141522207443\\ 96	0.743515493286708\\ 192	0.837429060339866\\ 240	0.859041451863658\\ };

\addplot [color=sceEleven,densely dotted,mark=\markSceEleven,mark options={solid},forget plot]
  table[row sep=crcr]{%
  24	0.494865879285207\\ 27	0.511384514554952\\ 30	0.529812546389712\\ 33	0.540103436213896\\ 36	0.547088964067227\\ 39	0.564042527238882\\ 42	0.574578960323017\\ 45	0.579648961632766\\ 48	0.590561706805632\\ 72	0.70773055994156\\ 96	0.770066022549458\\ 192	0.859993142217512\\ 240	0.883056523951399\\ };

\end{axis}

\node[draw, align=left, fill=white, anchor=south east, font=\small] at ($(current bounding box.south west)!.52!(current bounding box.south east)$) {solid: \nmimo \\ dashed: \gls{lsmimo} \\ dotted: \locPre};

\end{tikzpicture}%
  \caption{Jain's fairness index for 256 \gls{qam} and \merWf.}
  \label{fig:jain256merWf}
\end{figure}
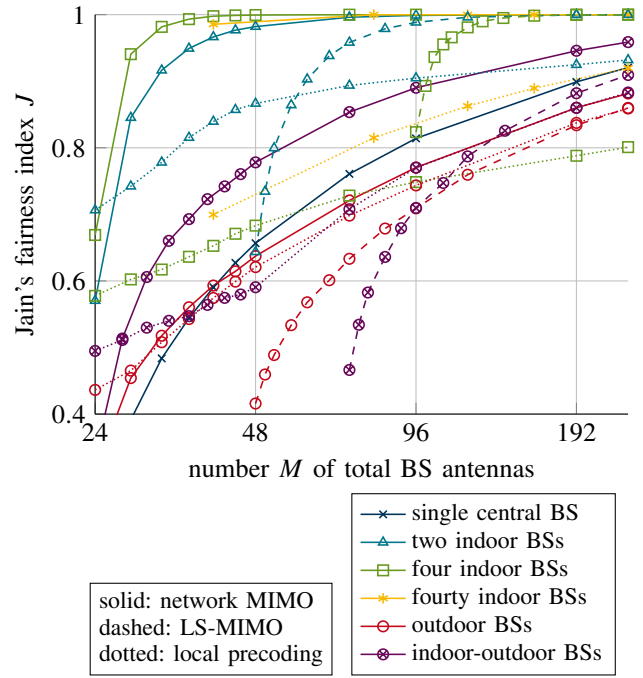

\figref{fig:jain256merWf} shows the simulated fairness indices.
For \nmimo and \gls{lsmimo} the \indtwoBS deployment, the \indfourBS deployment and the \indfourtyBS deployment approach perfect fairness indices of $1$ with an increasing number of \gls{bs} antennas. This is partly due to all \glspl{ue} being served with the maximal per-\gls{ue} \gls{se} of 256 \gls{qam}. With Gaussian modulation the trends of Jain's fairness index are similar, but no deployment achieves perfect fairness.
For \locPre the fairness indices are lower and they do not approach a fairness index of $1$ in the range of \gls{bs} antennas we consider.
The \indcenBS deployment, the \outBS deployment and the \inoutBS deployment do not approach a fairness index of $1$ with any transmission scheme in the range of \gls{bs} antennas, but the index increases with the number $M$ of \gls{bs} antennas.

We conclude that fairness increases with the number of \gls{bs} antennas, with the level of cooperation between \glspl{bs}, and with the distribution of \gls{bs} antennas (given some cooperation between \glspl{bs}).
Note that one can increase fairness by making it an objective while scheduling and allocating power.

\section{Noisy Channel Estimation}
\label{sec:noisyCSI}

So far we analyzed performance with perfect \gls{csi}. However, perfect \gls{csi} is usually not available.
We analyze the effect of estimation errors on the average \gls{se}.
We denote the channel coefficient with estimation error from the $m$-th antenna of the $i$-th \gls{bs} to the $k$-th \gls{ue} at subcarrier $f$ as
\begin{equation}
 \hat{h}_{i,k,m}^{(f)} = h_{i,k,m}^{(f)} + e_{i,k,m}^{(f)}
\end{equation}
where $h_{i,k,m}^{(f)}$ is the channel coefficient without error and $e_{i,k,m}^{(f)}$ is the estimation error. We model the estimation errors as independent and zero-mean proper complex Gaussian random variables. The estimation error of the channel between the $i$-th \gls{bs} and the $k$-th \gls{ue} is normalized such that its variance scales with the mean channel coefficient squared
\begin{equation}
 \E\left[\left|e_{i,k,m}^{(f)} \right|^2\right] = \E\left[\left|h_{i,k,m}^{(f)}\right|\right]^2 \sigma_E^2 
\end{equation}
where $\sigma_E^2 $ is the \gls{nmse}, and the expectation is over the \gls{bs} antennas and the subcarriers. This channel estimation error occurs, e.g., for channel prediction \cite{ofdmWorkshop}.
We determine the precoders based on the channel estimation with error.
For these precoders, intra-cell interference occurs due to the estimation error.

\begin{figure}[t!]
  \centering
%
%
%
\begin{tikzpicture}

\begin{axis}[%
width=\plotwidth,
height=\plotheight,
clip mode=individual,
scale only axis,
xmin=0,
xmax=50,
xtick={0,10,20,30,40,50},
xticklabels={$0$,$-10$,$-20$,$-30$,$-40$,$-50$},
xlabel={NMSE $\sigma_E^2$ [\SI{}{\dB}]},
xlabel style={inner ysep=\legOffSet,yshift=\legOffSetHalf,},
xmajorgrids,
ymin=30,
ymax=165,
ytick={40, 80, 120, 160},
ylabel={average sum \gls{se} $S$ [bit/s/Hz]},
ymajorgrids,
axis x line*=bottom,
axis y line*=left,
legend style={at={($(current bounding box.south west)!.55!(current bounding box.south east)$)},anchor=north west,draw=black,fill=white,legend cell align=left,font=\small,row sep=-1pt,},
every axis plot/.append style={semithick},
]

\addplot [color=sceTwo,dashed,mark=\markSceTwo,mark options={solid},forget plot]
  table[row sep=crcr]{%
0	105.105363342716\\
10	116.121697471036\\
20	116.991469870138\\
30	117.003523460534\\
40	117.001709698232\\
50	117.001467014157\\
};


\addplot [color=sceSeven,dashed,mark=\markSceSeven,mark options={solid},forget plot]
  table[row sep=crcr]{%
0	63.6345240267498\\
10	71.2687401140999\\
20	72.1744802890534\\
30	72.226021454644\\
40	72.2304888069745\\
50	72.2290964333626\\
};


\addplot [color=sceNine,dashed,mark=\markSceNine,mark options={solid},forget plot]
  table[row sep=crcr]{%
0	89.1429897728033\\
10	91.8395698539266\\
20	92.5507500386196\\
30	92.6280297262666\\
40	92.6347384321404\\
50	92.6353056315055\\
};


\addplot [color=sceTen,dashed,mark=\markSceTen,mark options={solid},forget plot]
  table[row sep=crcr]{%
0	56.7164625171314\\
10	66.4238527043829\\
20	67.666103678366\\
30	67.7912454190631\\
40	67.8040998850801\\
50	67.8056789873985\\
};


\addplot [color=sceEleven,dashed,mark=\markSceEleven,mark options={solid},forget plot]
  table[row sep=crcr]{%
0	52.0572400909519\\
10	61.4929124303897\\
20	62.398606088055\\
30	62.3380332436963\\
40	62.3217343765578\\
50	62.3187579666048\\
};


\addplot [color=sceTwo,solid,mark=\markSceTwo,mark options={solid}]
  table[row sep=crcr]{%
0	15.5192008237508\\
10	41.9221296295507\\
20	79.5539679397896\\
30	113.055480357336\\
40	134.849620135288\\
50	144.999015611254\\
};
\addlegendentry{\legSceTwo};

\addplot [color=sceSeven,solid,mark=\markSceSeven,mark options={solid}]
  table[row sep=crcr]{%
0	17.554947829328\\
10	45.58913342112\\
20	87.1847008317569\\
30	124.598842688282\\
40	147.073997730379\\
50	157.003975415506\\
};
\addlegendentry{\legSceSeven};

\addplot [color=sceNine,solid,mark=\markSceNine,mark options={solid}]
  table[row sep=crcr]{%
0	40.3390817728487\\
10	77.5311543902266\\
20	115.83483705209\\
30	136.911285871066\\
40	143.293944915108\\
50	144.440011326225\\
};
\addlegendentry{\legSceNine};

\addplot [color=sceTen,solid,mark=\markSceTen,mark options={solid}]
  table[row sep=crcr]{%
0	18.4488247608125\\
10	44.6322329314564\\
20	61.5309848290614\\
30	66.7111877178971\\
40	67.4423508759076\\
50	67.5171486963388\\
};
\addlegendentry{\legSceTen};

\addplot [color=sceEleven,solid,mark=\markSceEleven,mark options={solid}]
  table[row sep=crcr]{%
0	17.7695407442185\\
10	45.9179111354477\\
20	70.9166317166614\\
30	83.438805617882\\
40	88.693436315491\\
50	90.8370644289481\\
};
\addlegendentry{\legSceEleven};

\end{axis}

\node[draw, align=left, fill=white, anchor=south east, font=\small] at ($(current bounding box.south west)!.52!(current bounding box.south east)$) {dashed: \locPre \\ solid: \nmimo};

\end{tikzpicture}%
  \caption{Average \gls{se} with $48$ total \gls{bs} antennas (with $40$ total \gls{bs} antennas for the \indfourtyBS deployment) for a zero-mean Gaussian distributed channel estimation error.}
  \label{fig:avgSE256estErr}
\end{figure}
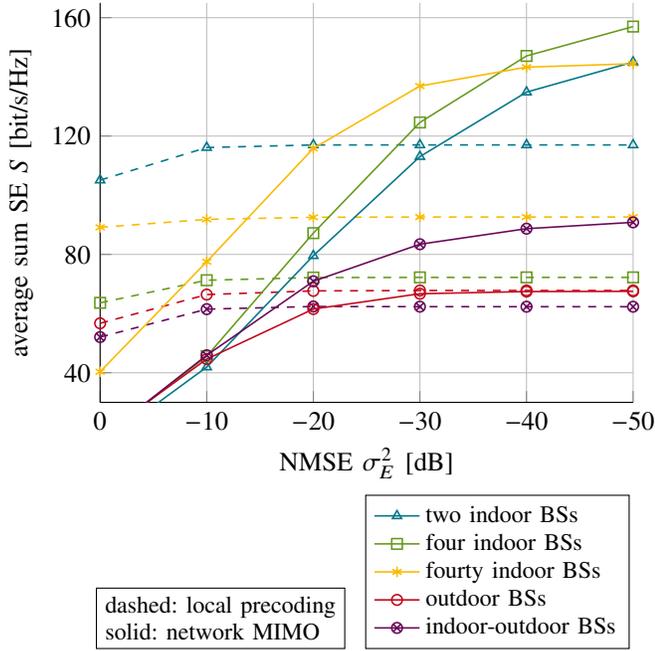

\figref{fig:avgSE256estErr} shows the average \glspl{se} versus \gls{nmse} for $48$ total transmit antennas, except for the \indfourtyBS deployment where we deploy only $40$ total \gls{bs} antennas. With \nmimo the performance of all deployments severely degrades with increasing \gls{nmse}. 
The \glspl{se} of \locPre are unaffected by low \gls{nmse} and degrade for high \gls{nmse} only. Inter-cell interference is always present for \locPre and dominates over the interference caused by channel estimation errors for most of the \gls{nmse} range. Hence the power allocation of \locPre is more robust to interference and \locPre outperforms \nmimo for a \gls{nmse} higher than \SIrange{-30}{-20}{\dB}. However, the performance of \nmimo with estimation errors can be improved, e.g., by making the power allocation more robust to the additional interference caused by estimation errors \cite{Yoo06}.


For more \gls{bs} antennas the trends and performance differences are similar.
We conclude that all deployments suffer from channel estimation noise, while some deployments are more sensitive. Good channel estimation is crucial to obtain the \mmimo and \nmimo benefits. However, more robust precoding techniques and power allocation schemes could improve performance in the presence of prediction errors.

\section{Conclusions}
\label{sec:conc}

We compared the performance of six different deployments and different levels of cooperation in the \gls{3gpp} indoor office scenario.
Cooperation between \glspl{bs} provides gains as compared to no cooperation, which become larger as the level of cooperation increases. The same performance as a single massive \gls{mimo} \gls{bs} is achieved by distributed \glspl{bs} with cooperation and fewer antennas. The costs of antenna elements can be traded off with the costs for backhaul capability to achieve the same performance. A ratio of twice as many \gls{bs} antennas as served \glspl{ue} offers many of the massive \gls{mimo} benefits. User fairness and \gls{se} close to capacity are achieved with a simple transmission scheme. Accurate channel estimation is necessary to obtain the \mmimo and cooperation benefits.

A \gls{se} of \SI{100}{\bit\per\second\per\hertz} without considering overhead is achievable with $192$ antennas using \locPre, and less than $28$ antennas using two indoor \glspl{bs} with \nmimo. Considering an overhead of \SI{50}{\percent}, the required bandwidth to achieve the goals of the \acrshort{metis} project \cite{metis_d11} is:
\begin{itemize}
 \item For the TC1 virtual reality office:
 \begin{equation}
  \frac{\SI{0.1}{\giga\bit\per\second\per\meter\squared} \cdot \SI{5000}{\meter\squared} }{ \SI{50}{\bit\per\second\per\hertz} } = \SI{10}{\giga\hertz}.
 \end{equation}
  More \gls{ue} antennas, more base stations, or larger \gls{qam} constellations could reduce the required bandwidth.
 \item For the TC2 dense urban information society:
 \begin{equation}
  \frac{\SI{0.7}{\mega\bit\per\second\per\meter\squared} \cdot \SI{5000}{\meter\squared} }{ \SI{50}{\bit\per\second\per\hertz} } = \SI{70}{\mega\hertz}.
 \end{equation}
  This performance is achievable with single antenna \glspl{ue}, few \glspl{bs}, and 256 \gls{qam} within a reasonable bandwidth.
\end{itemize}

\section*{Acknowledgment}
S. Dierks and G. Kramer were supported in part by an Alexander von Humboldt Professorship endowed by the German Federal Ministry of Education and Research.

\ifCLASSOPTIONcaptionsoff
  \newpage
\fi



\bibliographystyle{IEEEtran}
%
\bibliography{refs}

\begin{thebibliography}{10}
\providecommand{\url}[1]{#1}
\csname url@samestyle\endcsname
\providecommand{\newblock}{\relax}
\providecommand{\bibinfo}[2]{#2}
\providecommand{\BIBentrySTDinterwordspacing}{\spaceskip=0pt\relax}
\providecommand{\BIBentryALTinterwordstretchfactor}{4}
\providecommand{\BIBentryALTinterwordspacing}{\spaceskip=\fontdimen2\font plus
\BIBentryALTinterwordstretchfactor\fontdimen3\font minus
  \fontdimen4\font\relax}
\providecommand{\BIBforeignlanguage}[2]{{%
\expandafter\ifx\csname l@#1\endcsname\relax
\typeout{** WARNING: IEEEtran.bst: No hyphenation pattern has been}%
\typeout{** loaded for the language `#1'. Using the pattern for}%
\typeout{** the default language instead.}%
\else
\language=\csname l@#1\endcsname
\fi
#2}}
\providecommand{\BIBdecl}{\relax}
\BIBdecl

\bibitem{metis_d11}
{Mobile and Wireless Communications Enablers for the Twenty-Twenty Information
  Society (METIS)}, ``Deliverable {D1.1} - scenarios, requirements and {KPIs}
  for {5G} mobile and wireless system,'' Tech. Rep., Apr. 2013.

\bibitem{Marzetta_mMIMO2010}
T.~Marzetta, ``Noncooperative cellular wireless with unlimited numbers of base
  station antennas,'' \emph{IEEE Trans. Wireless Commun.}, vol.~9, no.~11, pp.
  3590--3600, Nov. 2010.

\bibitem{Larsson2014}
E.~Larsson, O.~Edfors, F.~Tufvesson, and T.~Marzetta, ``Massive {MIMO} for next
  generation wireless systems,'' \emph{IEEE Commun. Mag.}, vol.~52, no.~2, pp.
  186--195, Feb. 2014.

\bibitem{Rusek13}
F.~Rusek, D.~Persson, B.~K. Lau, E.~G. Larsson, T.~L. Marzetta, O.~Edfors, and
  F.~Tufvesson, ``Scaling up {MIMO}: Opportunities and challenges with very
  large arrays,'' \emph{IEEE Signal Process. Mag.}, vol.~30, no.~1, pp. 40--60,
  Jan. 2013.

\bibitem{Bjoernson16}
E.~Bj\"{o}rnson, E.~G. Larsson, and T.~L. Marzetta, ``Massive {MIMO}: Ten myths
  and one critical question,'' \emph{IEEE Commun. Mag.}, vol.~54, no.~2, pp.
  114--123, Feb. 2016.

\bibitem{bjoernson14}
E.~Bj\"{o}rnson, M.~Bengtsson, and B.~Ottersten, ``Optimal multiuser transmit
  beamforming: A difficult problem with a simple solution structure [lecture
  notes],'' \emph{IEEE Signal Process. Mag.}, vol.~31, no.~4, pp. 142--148,
  Jul. 2014.

\bibitem{Costa83}
M.~Costa, ``Writing on dirty paper,'' \emph{IEEE Trans. Inf. Theory}, vol.~29,
  no.~3, pp. 439--441, May 1983.

\bibitem{Femtocells.2010}
J.~Zhang and G.~de~la Roche, Eds., \emph{Femtocells: Technologies and
  Deployment}.\hskip 1em plus 0.5em minus 0.4em\relax John Wiley \& Sons, 2013.

\bibitem{3GPP_TR36_814.2010}
3GPP, ``{TR36.814} - further advancements for {E-UTRA} physical layer
  aspects,'' Tech. Rep. v9.0.0, Mar. 2010.

\bibitem{jainsindex}
R.~Jain, D.-M. Chiu, and W.~R. Hawe, ``A quantitative measure of fairness and
  discrimination for resource allocation in shared computer systems,'' DEC
  Research Report, Tech. Rep. 301, 1984.

\bibitem{ofdmWorkshop}
S.~Dierks, M.~Amin, W.~Zirwas, M.~Haardt, and B.~Panzner, ``The benefit of
  cooperation in the context of massive {MIMO},'' in \emph{18th Int. OFDM
  Workshop (InOWoS)}, Aug. 2014, pp. 1--8.

\bibitem{globeComMMIMOWS}
B.~Panzner, W.~Zirwas, S.~Dierks, M.~Lauridsen, P.~Mogensen, K.~Pajukoski, and
  D.~Miao, ``Deployment and implementation strategies for massive {MIMO} in
  {5G},'' in \emph{IEEE Global Telecommun. Conf. (Globecom) Workshop Massive
  MIMO}, Dec. 2014.

\bibitem{eusipco}
S.~Dierks, W.~Zirwas, M.~J\"{a}ger, B.~Panzner, and G.~Kramer, ``{MIMO} and
  massive {MIMO} - analysis for a local area scenario,'' in \emph{23nd European
  Signal Processing Conf. (EUSIPCO)}, Sep. 2015, pp. 2496--2500.

\bibitem{LozanoJul06}
A.~Lozano, A.~M. Tulino, and S.~Verdu, ``Optimum power allocation for parallel
  {Gaussian} channels with arbitrary input distributions,'' \emph{IEEE Trans.
  Inf. Theory}, vol.~52, no.~7, pp. 3033--3051, Jul. 2006.

\bibitem{IST-WINNERIIDeliverable1.1.22008a}
IST, ``D1.1.2 - {WINNER II} channel models,'' Tech. Rep. v1.2, 2008.

\bibitem{Boudreau09}
G.~Boudreau, J.~Panicker, N.~Guo, R.~Chang, N.~Wang, and S.~Vrzic,
  ``Interference coordination and cancellation for {4G} networks,'' \emph{IEEE
  Commun. Mag.}, vol.~47, no.~4, pp. 74--81, Apr. 2009.

\bibitem{Lee12}
D.~Lee, H.~Seo, B.~Clerckx, E.~Hardouin, D.~Mazzarese, S.~Nagata, and
  K.~Sayana, ``Coordinated multipoint transmission and reception in
  {LTE}-advanced: Deployment scenarios and operational challenges,'' \emph{IEEE
  Commun. Mag.}, vol.~50, no.~2, pp. 148--155, Feb. 2012.

\bibitem{CIT069}
\BIBentryALTinterwordspacing
E.~Bj\"{o}rnson and E.~Jorswieck, ``Optimal resource allocation in coordinated
  multi-cell systems,'' \emph{Foundations and Trends in Communications and
  Information Theory}, vol.~9, no. 2-3, pp. 113--381, 2013. [Online].
  Available: \url{http://dx.doi.org/10.1561/0100000069}
\BIBentrySTDinterwordspacing

\bibitem{Wiesel2008}
A.~Wiesel, Y.~Eldar, and S.~Shamai, ``Zero-forcing precoding and generalized
  inverses,'' \emph{IEEE Trans. Signal Proc.}, vol.~56, no.~9, pp. 4409--4418,
  Sep. 2008.

\bibitem{Zhang2008}
X.~Zhang and J.~Lee, ``Low complexity {MIMO} scheduling with channel
  decomposition using capacity upperbound,'' \emph{IEEE Trans. Commun.},
  vol.~56, no.~6, pp. 871--876, Jun. 2008.

\bibitem{Gesbert10}
D.~Gesbert, S.~Hanly, H.~Huang, S.~S. Shitz, O.~Simeone, and W.~Yu,
  ``Multi-cell {MIMO} cooperative networks: A new look at interference,''
  \emph{IEEE J. Sel. Areas Commun.}, vol.~28, no.~9, pp. 1380--1408, Dec. 2010.

\bibitem{Hosseini14}
K.~Hosseini, W.~Yu, and R.~S. Adve, ``Large-scale {MIMO} versus network {MIMO}
  for multicell interference mitigation,'' \emph{IEEE J. Sel. Topics Signal
  Process.}, vol.~8, no.~5, pp. 930--941, Oct. 2014.

\bibitem{Zhang10}
R.~Zhang, ``Cooperative multi-cell block diagonalization with per-base-station
  power constraints,'' \emph{IEEE J. Sel. Areas Commun.}, vol.~28, no.~9, pp.
  1435--1445, Dec. 2010.

\bibitem{Shi08}
S.~Shi, M.~Schubert, N.~Vucic, and H.~Boche, ``{MMSE} optimization with
  per-base-station power constraints for network {MIMO} systems,'' in
  \emph{IEEE Int. Conf. Communications (ICC)}, May 2008, pp. 4106--4110.

\bibitem{zhang2004capacity}
H.~Zhang and H.~Dai, ``On the capacity of distributed {MIMO} systems,'' in
  \emph{38th Conf. Inform. Sciences and Systems (CISS)}, 2004.

\bibitem{pcell15}
A.~Forenza, S.~Perlman, F.~Saibi, M.~D. Dio, R.~van~der Laan, and G.~Caire,
  ``Achieving large multiplexing gain in distributed antenna systems via
  cooperation with {pCell} technology,'' in \emph{49th Asilomar Conf. Signals,
  Systems and Computers (ACSSC)}, Nov. 2015, pp. 286--293.

\bibitem{quadriga}
S.~Jaeckel, L.~Raschkowski, K.~B\"{o}rner, L.~Thiele, F.~Burkhardt, and
  E.~Eberlein, ``{QuaDRiGa}-quasi determinsitic radio channel generator, user
  manual and documentation,'' Fraunhofer Heinrich Hertz Institute, Tech. Rep.
  v1.4.1-551, 2016.

\bibitem{winnerPlusD5.3}
CELTIC, ``D5.3: {WINNER+} final channel models,'' Tech. Rep. v1.0, 2010.

\bibitem{Ungerboeck82}
G.~Ungerboeck, ``Channel coding with multilevel/phase signals,'' \emph{IEEE
  Trans. Inf. Theory}, vol.~28, no.~1, pp. 55--67, Jan. 1982.

\bibitem{Caire03}
G.~Caire and S.~Shamai, ``On the achievable throughput of a multiantenna
  {Gaussian} broadcast channel,'' \emph{IEEE Trans. Inf. Theory}, vol.~49,
  no.~7, pp. 1691--1706, Jul. 2003.

\bibitem{ViswanathTse03}
P.~Viswanath and D.~N.~C. Tse, ``Sum capacity of the vector {Gaussian}
  broadcast channel and uplink-downlink duality,'' \emph{IEEE Trans. Inf.
  Theory}, vol.~49, no.~8, pp. 1912--1921, Aug. 2003.

\bibitem{VishwJind03}
S.~Vishwanath, N.~Jindal, and A.~Goldsmith, ``Duality, achievable rates, and
  sum-rate capacity of {Gaussian} {MIMO} broadcast channels,'' \emph{IEEE
  Trans. Inf. Theory}, vol.~49, no.~10, pp. 2658--2668, Oct. 2003.

\bibitem{Yu04}
W.~Yu and J.~M. Cioffi, ``Sum capacity of {Gaussian} vector broadcast
  channels,'' \emph{IEEE Trans. Inf. Theory}, vol.~50, no.~9, pp. 1875--1892,
  Sep. 2004.

\bibitem{jindal05}
N.~Jindal, W.~Rhee, S.~Vishwanath, S.~Jafar, and A.~Goldsmith, ``Sum power
  iterative water-filling for multi-antenna {G}aussian broadcast channels,''
  \emph{IEEE Trans. Inf. Theory}, vol.~51, no.~4, pp. 1570--1580, Apr. 2005.

\bibitem{Yoo06}
T.~Yoo and A.~Goldsmith, ``Capacity and power allocation for fading {MIMO}
  channels with channel estimation error,'' \emph{IEEE Trans. Inf. Theory},
  vol.~52, no.~5, pp. 2203--2214, May 2006.

\end{thebibliography}

%







\begin{IEEEbiography}[{\includegraphics[width=1in,height=1.25in,clip,keepaspectratio]{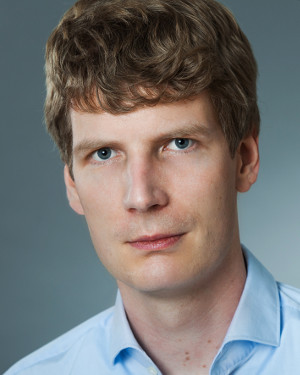}}]{Stefan Dierks}
received the B.Sc. and the M.Sc. equivalent Diplom-Ingenieur degree in Electrical Engineering from the Technical University of Munich (TUM) in 2009 and 2011, respectively. Since 2011 he is working towards his doctoral degree at the Chair for Communications Engineering at TUM under the supervision of Prof. Kramer. His research interests include 5G, massive MIMO, cooperative communications, interference alignment, and EIRP.
\end{IEEEbiography}

\begin{IEEEbiography}[{\includegraphics[width=1in,height=1.25in,clip,keepaspectratio]{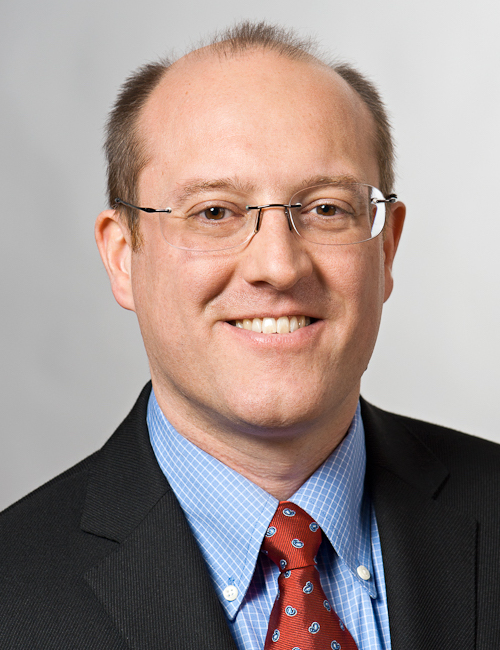}}]{Gerhard Kramer}
is Alexander von Humboldt Professor and Chair of Communications Engineering at the Technical University of Munich (TUM). He received the B.Sc. and M.Sc. degrees in electrical engineering from the University of Manitoba, Canada, in 1991 and 1992, respectively, and the Dr. sc. techn. degree from the ETH Zurich, Switzerland, in 1998. From 1998 to 2000, he was with Endora Tech AG in Basel, Switzerland, and from 2000 to 2008 he was with the Math Center at Bell Labs in Murray Hill, NJ. He joined the University of Southern California (USC), Los Angeles, CA, as a Professor of Electrical Engineering in 2009. He joined TUM in 2010.
Gerhard Kramer's research interests are primarily in information theory and communications theory, with applications to wireless, copper, and optical fiber networks. 
\end{IEEEbiography}

\begin{IEEEbiography}[{\includegraphics[width=1in,height=1.25in,clip,keepaspectratio]{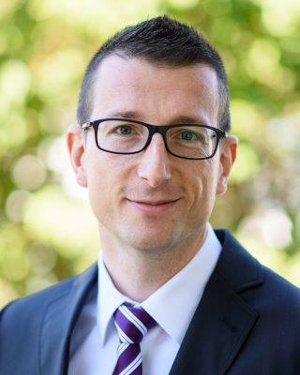}}]{Berthold Panzner}
received his M.Sc. degree in telecommunications in 2007 from Link\"{o}pings University in Sweden and the PhD degree in electrical engineering from Otto-von-Guericke University Magdeburg in Germany in 2012. He received two best paper awards at the Antenna and Propagation Symposium 2008 and the International Radar Symposium 2012 respectively for his research work on synthetic aperture radar techniques for ground penetrating radar. He joint Nokia Siemens Networks in 2013, where he was involved in the design of physical layer aspects and massive MIMO for 5G. He is currently working at Nokia Networks as Solution Architect for IoT connectivity solutions.
\end{IEEEbiography}

\begin{IEEEbiography}[{\includegraphics[width=1in,height=1.25in,clip,keepaspectratio]{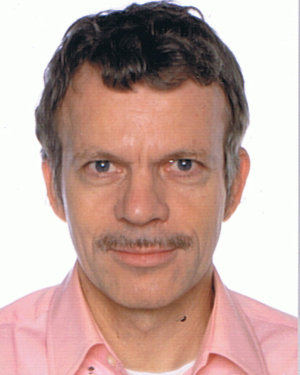}}]{Wolfgang Zirwas}
received his diploma degree in communication technologies in 1987 from Technical University of Munich (TUM). He started his work at the Siemens Munich central research lab for communication technologies with a focus on high frequency and high data rate TDM fiber systems for data rates up to 40 Gbit/s. He participated in several German and EU funded projects like COVERAGE, WINNER, Artist4G, METIS or Fantstic5G. Currently, he is at the end to end Lab of Nokia Bell Labs in Munich investigating 5G mobile radio technologies. 
\end{IEEEbiography}

\vfill

\end{document}